\documentclass[%
 reprint,
superscriptaddress,
preprintnumbers,
 amsmath,amssymb,
 aps,
 prc,
]{revtex4-2}

\usepackage[english]{babel}
\usepackage{graphicx}  
\usepackage{dcolumn}   
\usepackage{bm}        
\usepackage{grffile}
\usepackage{xspace}
\usepackage{ifthen}
\usepackage{xifthen}
\usepackage{xparse}
\usepackage{afterpage}
\usepackage{delimset}
\usepackage{stmaryrd}
\usepackage{xcolor}
\usepackage{comment}
\usepackage{physics}
\usepackage{isotope}
\usepackage{multirow}
\usepackage[detect-all]{siunitx}
\usepackage{hyperref}

\newcommand{\gras}[1]{\boldsymbol{#1}}

\newcommand{\qvec}         {\vb{q}}

\newcommand{\prob}  [2][]  {%
  \ensuremath{%
    \mathbb{P}\ifthenelse{\isempty{#1}}%
      {\left({#2}\right)}%
      {\left({#2}\,\middle|\,{#1}\right)}%
  }\xspace%
}

\newcommand{\AR}           {{A_{\rm R}}}

\newcommand{\sigmaARes}    {\ensuremath{\sigma_{\rm A}}}
\newcommand{\Hnucl}        {\ensuremath{\ope{H}}}
\newcommand{\Jz}           {\ensuremath{\ope{J}_{\rm z}}}

\newcommand{\qneck}        {q_{\rm N}}
\newcommand{\qnecksciss}   {q_{\rm N}^{\rm sciss}}

\newcommand{\quadru}       {q_{\rm 20}}
\newcommand{\octu}         {q_{\rm 30}}

\newcommand{\EFI}         {E_{\rm FI}}
\newcommand{\EA}          {E_{\rm A}}
\newcommand{\EB}          {E_{\rm B}}

\newcommand{\En}          {E_{\rm n}}
\NewDocumentCommand{\EBbind} {d<>}     {{\ensuremath{E_{\rm bind.}\IfValueT{#1}{^{(#1)}}}}}

\newcommand{\ope}    [1]    {\ensuremath{\hat{#1}}}

\NewDocumentCommand{\GenOpColl}   {m m d<>}   {\ensuremath{{\mathnormal{#2}_{\rm coll.}\IfValueTF{#3}{^{(#3)\IfBooleanT{#1}{\star}}}{\IfBooleanT{#1}{^{\star}}}}}}
\NewDocumentCommand{\Hcoll}       {d<> s}     {{\ensuremath{\GenOpColl{#2}{H}<#1>}}}
\NewDocumentCommand{\Acoll}       {d<> s}     {{\ensuremath{\GenOpColl{#2}{A}<#1>}}}
\NewDocumentCommand{\GCMstate}    {d<> e{_}}  {{\ensuremath{g\IfValueT{#1}{^{(#1)}}\IfValueT{#2}{_{#2}}}}}
\newcommand{\GCMstateInit}{g_{n}^{(K)}}
\NewDocumentCommand{\GCMEne}      {d<> e{_}}  {{\ensuremath{E\IfValueT{#1}{^{(#1)}}\IfValueT{#2}{_{#2}}}}}
\NewDocumentCommand{\GCMEneAvg}   {d<>}       {{\ensuremath{\bar{E}\IfValueT{#1}{^{(#1)}}}}}
\newcommand{\GCMEneInit}      {E_0}
\newcommand{\GCMSigInit}      {\sigma}

\newcommand{\ladder}[4]    {%
  \def\tempid{#1}%
  \def\tempac{#2}%
  \ifx\tempid\empty%
    {#3}_{#4}\ifx#2c^{\dagger}\fi%
  \else%
    {#3}^{(#1)\ifx#2c{\dagger}\fi}_{#4}%
  \fi%
}
\newcommand{\ladderc}[4]    {%
  \def\tempid{#1}%
  \def\tempac{#2}%
  \ifx\tempid\empty%
    {#3}({#4})\ifx#2c^{\dagger}\fi%
  \else%
    {#3}^{(#1)\ifx#2c{\dagger}\fi}({#4})%
  \fi%
}

\newcommand{\figref}  [1]{Fig.~\ref{#1}}
\newcommand{\secref}  [1]{Sec.~\ref{#1}}

\begin{document}

\title{Microscopic Calculation of Fission Product Yields for Odd-Mass Nuclei}
\author{N. Schunck}
\email{schunck1@llnl.gov}
\affiliation{
Nuclear and Chemical Sciences Division, Lawrence Livermore National Laboratory, Livermore, California 94551, USA
}

\author{M. Verriere}
\affiliation{
Nuclear and Chemical Sciences Division, Lawrence Livermore National Laboratory, Livermore, California 94551, USA
}

\author{G. Potel Aguilar}
\affiliation{
Nuclear and Chemical Sciences Division, Lawrence Livermore National Laboratory, Livermore, California 94551, USA
}

\author{R. C. Malone}
\affiliation{
Nuclear and Chemical Sciences Division, Lawrence Livermore National Laboratory, Livermore, California 94551, USA
}

\author{J. A. Silano}
\affiliation{
Nuclear and Chemical Sciences Division, Lawrence Livermore National Laboratory, Livermore, California 94551, USA
}

\author{A. P. D. Ramirez}
\affiliation{
Nuclear and Chemical Sciences Division, Lawrence Livermore National Laboratory, Livermore, California 94551, USA
}

\author{A. P. Tonchev}
\affiliation{
Nuclear and Chemical Sciences Division, Lawrence Livermore National Laboratory, Livermore, California 94551, USA
}
\affiliation{
Department of Physics, Duke University, Durham, NC 27708-0308, USA
}

\date{\today}
\preprint{LLNL-JRNL-XXXXXX}

\begin{abstract}
Fission data are essential inputs to reaction networks involved in 
nucleosynthesis simulations and nuclear forensics. In such applications as well
as in the description of multi-chance fission, the characteristics of fission 
for odd-mass nuclei are just as important as those for even-even nuclei. The 
fission theories that aim at explicitly describing fission dynamics are 
typically based on some variant of the nuclear mean-field theory. In such 
cases, the treatment of systems with an odd number of particles is markedly
more involved, both formally and computationally. In this article, we use the
blocking prescription of the Hartree-Fock-Bogoliubov theory with Skyrme energy
functionals to compute the deformation properties of odd-mass uranium isotopes.
We show that the resulting fission fragment distributions depend quite
significantly on the spin of the odd neutron. By direct calculation of the spin
distribution of the fissioning nucleus, we propose a methodology to rigorously
predict the charge and mass distributions in odd-mass nuclei.
\end{abstract}


\maketitle

\section{Introduction}
\label{sec:intro}

The theory of nuclear fission has a long and rich history 
\cite{schunck2022theory} and yet, it is still undergoing a spectacular 
renaissance \cite{bender2020future}. Thanks to the continuous increase in 
computing capabilities, microscopic methods based on nuclear density functional 
theory (DFT) have become very competitive with the phenomenological models that 
were prevalent until now \cite{schunck2016microscopic}. The application of 
these techniques has given truly novel insights into the fission process, such 
as unveiling the role of shell effects in setting the dominant fission modes 
\cite{scamps2018impact}, analyzing the dissipative nature of the fission 
process \cite{scamps2015superfluid,tanimura2015collective,bulgac2016induced,
bulgac2019fission}, and predicting the spin of fission fragments 
\cite{bulgac2021fission,marevic2021angular,bulgac2022fragment}. These recent 
developments are all the more important as simultaneous progress in simulations 
of nucleosynthesis have created a strong demand for predictive and complete 
models of fission applicable across the entire chart of isotopes 
\cite{martinez-pinedo2007role,thielemann2011what,mumpower2016impact}. A more 
predictive model of fission may also be key to understanding the nuclear 
reactor anti-neutrino anomaly \cite{sonzogni2017dissecting,
schmidt2021extensive}. Such applications make it especially important to build 
comprehensive models of fission that can describe the entire chain of events 
occuring in the process, from the formation of the compound nucleus to the 
$\beta$ decay and delayed emission of the fission products 
\cite{lovell2021extension,kawano2021influence,talou2021fission}.

Until now, the majority of fission studies, whether based on DFT or 
phenomenological mean-field models, have been restricted to even-even 
fissioning nuclei. There are a few notable exceptions: in 
Refs.~\cite{hock2013fission,koh2017fission}, fission barrier heights of 
$^{235}$U and $^{239}$Pu were computed in the Skyrme Hartree-Fock theory with 
pairing correlations treated at the BCS approximation, including a full 
treatment of the time-odd terms for several different configurations of the odd 
neutron. In Ref.~\cite{perez-martin2009fission}, the one-dimensional fission 
path and spontaneous fission half-life of $^{235}$U for two different values of 
the spin projection $K=1/2$ and $K=7/2$ were computed at the 
Hartree-Fock-Bogoliubov (HFB) approximation with the Gogny force. This analysis 
was extended in Ref.~\cite{rodriguez-guzman2017microscopic}, where similar 
calculations were performed for the uranium and plutonium isotopic chains. Even 
though the details of these studies differ, they all highlighted the fact that 
fission barrier heights, and by extension, the potential energy surface, vary 
substantially with the configuration occupied by the odd particle. While this 
has an obvious impact on calculated spontaneous fission half-lives, which are 
extremely sensitive to the shape and height of the fission barrier, one should 
also expect an effect on fission fragment distributions. Moreover, the spin 
dependence of the potential energy surface in odd-mass nuclei has an 
interesting consequence for neutron-induced fission since upon formation the 
compound nucleus acquires a sizable spin distribution. The characteristics of 
the entrance channel should therefore have a visible impact on the distribution 
of fission fragments.

The goal of this paper is to outline a theoretical framework based on the HFB 
theory with blocking and the time-dependent generator coordinate method with 
the Gaussian overlap approximation (TDGCM+GOA) to compute the charge and mass 
distributions of fission fragments for an odd-mass compound nucleus. We confirm 
the important impact of the blocked configurations on nuclear deformation 
properties. By computing two-dimensional potential energy surfaces, we give the 
first microscopic calculation of fission fragment distributions for different 
blocked configurations. Finally, we use the coupled channel reaction formalism 
to include information about the entrance channel in the determination of the 
fission fragment distributions.

Section \ref{sec:theory} gives an overview of the theoretical framework. Most 
of it is well known and presented in textbooks \cite{schunck2019energy}. The 
one exception is the generalization of the formula for the collective inertia 
to the case of an odd nucleus. In Section \ref{sec:static}, we summarize the 
results of static HFB calculations, both one-dimensional fission paths and 
two-dimensional potential energy surfaces. Section \ref{sec:fpy} discusses the 
methodology adopted to incorporate calculated spin distributions of the 
compound nucleus into predictions of fission fragment distributions and shows 
results for uranium isotopes.

\section{Theoretical Framework}
\label{sec:theory}

In this article, we focus on calculations of fission product yields from 
neutron-induced fission. After absorbing the incident neutron, the resulting 
compound nucleus can decay through different channels, either by emitting 
particles (primarily neutrons or $\gamma$ rays) or by fissioning. From a 
theoretical point of view, fission is described as a large-amplitude collective 
motion that drives the nucleus from a near-spherical shape to the scission 
point. Typically each fission event leads to a pair of fission fragments; 
ternary fission will not be considered here.We distinguish between the yields
at two different times following Ref.~\cite{mills1995fission}. The fission 
fragment charge and mass distributions that one would obtain immediately after 
scission will be referred to as {\it primary fission fragment distributions}. 
The term primary fission fragments means that the nuclear species formed have 
yet to emit any prompt particles. One can define primary charge $Y(Z)$, mass 
$Y(A)$, or isotopic $Y(Z,A)$ distributions.

The prompt emission of neutrons and $\gamma$ rays from the fission fragments 
changes the relative abundance of each isotope. At the end of this prompt 
deexcitation phase, which typically takes of the order of $10^{-16}$ s after 
scission, the new distributions of fission fragments are called the {\it 
independent yields}. Again, one distinguishes between independent charge 
$Y_{\rm ind.}(Z)$, mass $Y_{\rm ind.}(A)$, and isotopic $Y_{\rm ind.}(Z,A)$ 
yields.

\subsection{Treatment of the Entrance Channel}
\label{subsec:entrance}

For the heavy systems we are addressing in this work, the number of degrees of 
freedom involved in a neutron-induced reaction is very large, as attested by 
the high level density at the relevant excitation energies (typically of the 
order of $10^{6}$ MeV$^{{-1}}$ around the neutron separation energy). In this
regime, a statistical description of the process is known to work well 
\cite{feshbach1980statistical,koning2008talys1,herman2007empire,
kawano2010monte,ormand2021monte}, and the states populated in the reaction are 
described in terms of compound nucleus formation. In particular, Bohr's 
hypothesis is usually applied, according to which the way a compound nucleus 
decays is independent of how it was formed. Aside from the explicit 
consideration of small deviations from this hypothesis in terms of the 
so-called width fluctuations that correlate entrance and exit channels, there i
s an important caveat: the energy, angular momentum, and parity of the entrance 
channel are exactly preserved in the exit channel. Since the decay branching
ratios corresponding to the different decay modes depend strongly on these 
conserved quantities, it is essential to predict the population of the 
compound nucleus states in terms of the energy, spin, and parity distributions a
s the neutron is absorbed.

The absorption process is described within a direct reaction scheme in terms of 
a coupled-channels reaction formalism. Under the assumption that the target 
nucleus is described by a rigid rotor Hamiltonian (e.g., Ref.~\cite{Bohr:75}), 
the incident neutron is coupled to the intrinsic structure of the compound 
nucleus through the imaginary part of the optical potential, as well as through 
the direct excitation of the members of the ground-state rotational band. The 
wavefunction of the composite system formed by the incident neutron and the 
target is expanded in terms of the states $\psi_{i}(\xi)$ of the target 
ground-state rotational band (for simplicity, we omit angular momentum 
considerations; see Ref.~\cite{Tamura:65} for more details),
\begin{align}
\Psi(r,\xi)=\sum_{i}\phi_{i}(r)\psi_{i}(\xi),
\end{align} 
where $r$ is the neutron-target coordinate, and $\xi$ denotes the set of 
coordinates associated with the target. The set of coupled differential 
equations obeyed by the neutron wavefunctions $\phi_{i}(r)$ can be obtained by 
projecting the many-body Schr\"odinger equation on the target rotational states 
\cite{Tamura:65},
\begin{align}\label{eq_cc1}
\nonumber \left(\frac{\hbar^{2}}{2m}\frac{d^{2}}{dr^2}+\right.&
\left.\frac{\hbar^2 \ell_{i}(\ell_{i}+1)}{2mr^{2}}+U(r)-E_{i}\right)\phi_{i}(r)\\
&=-\sum_{j\neq i}V_{ij}(r;\gras{\beta})\phi_{j}(r),
\end{align} 
where $U(r)$ is a complex optical potential, and the subindex $i$ designates 
both the states of the rotational band and the set of quantum numbers of the 
neutron wavefunction. For a rotational nucleus, the coupling potentials 
$V_{ij}(r;\gras{\beta})$ depend on the deformation parameters $\gras{\beta} 
\equiv\{\beta_{2},\beta_{4},\beta_{6}\}$ of the mean-field potential in the 
target. Since we are using the coupling scheme developed in 
Ref.~\cite{Soukhovitskii:16} restricted to transitions within the ground-state 
rotational band, thus neglecting transitions between bands, we will only 
consider here deformations with even multipolarity. These potentials are 
typically defined through the following multipole expansion,
\begin{align}\label{eq_cc2}
V_{ij}(r;\beta)=\sum_{\lambda} v_{\lambda}(r;\beta_{\lambda})\,B_{\lambda}(i,j)\,A(i,j),
\end{align}
where $B_{\lambda}(i,j)$ and $A(i,j)$ are geometrical coefficients depending on 
the spins of the states $i$ and $j$ \cite{Tamura:65}. The functions 
$v_{\lambda}(r;\beta_{\lambda})$ are the coefficients of the multipole 
expansion of a deformed Woods-Saxon potential with standard real and imaginary 
surface terms, as well as a real spin-orbit term; for details, see 
Ref.~\cite{Soukhovitskii:16}. The geometry and energy dependence of this 
potential, as well as the deformation parameters $\beta$, have been fitted to 
reproduce neutron elastic scattering and total reaction cross sections in 
actinides \cite{Soukhovitskii:16}.

The set of coupled differential equations (\ref{eq_cc1}) is solved  to obtain 
the scattering matrix $S_{j,\pi}$ for the incident neutron, from which the 
transmission coefficient $T_{j,\pi}$ associated with the formation of the 
compound nucleus for each spin $j$ and parity $\pi$ can be obtained according 
to
\begin{align}\label{eq_cc3}
T_{j,\pi}=1-\left| S_{j,\pi} \right|^{2}.
\end{align}
In Sec.~\ref{subsec:fpy_ini} this transmission coefficient will be taken as the 
probability $P_{\rm th.}(J^{\pi})$ of having the compound nucleus in a state 
with spin $J$ and parity $\pi$.

\subsection{Large-Amplitude Collective Motion for Odd Nuclei}
\label{subsec:lacm}

One of the main goals of this work is to determine the primary fission fragment
distributions $Y(Z)$ and $Y(A)$ of an odd-mass fissioning nucleus. To this end,
we work within the global framework of the energy density functional (EDF)
theory \cite{schunck2019energy}. Fission fragment distributions are computed in 
a three-step process: (i) the potential energy surface (PES) of the nucleus is 
computed in a small space of collective variables within the static HFB theory; 
(ii) the time-evolution of a collective wavepacket on this PES is simulated 
with the time-dependent generator coordinate method under the Gaussian overlap 
approximation \cite{verriere2020timedependent}; (iii) the actual fission 
fragment charge and mass distributions are extracted by computing the flux of 
the collective wave packet across scission. This approach was invented in the
1980s at CEA Bruy\`eres-le-Ch\^atel \cite{berger1984microscopic,
berger1986quantum,berger1991timedependent} with early applications in the 2000s 
\cite{goutte2004mass,goutte2005microscopic,younes2012fragment} and it is 
presented in great detail in Refs.~\cite{regnier2016fission,regnier2016felix1,
younes2019microscopic,verriere2021microscopic}. In the following, we only 
describe the extension of this formalism to odd-mass nuclei.

\subsubsection{Blocking Prescription}
\label{subsubsec:blocking}

Nuclei with odd numbers of particles are computed at the HFB approximation with
the blocking prescription \cite{mang1975selfconsistent,faessler1981description,
duguet2001pairing}. The ansatz for the many-body state thus reads
\begin{equation}
\ket{\Phi} = \beta_{\alpha}^{\dagger}\prod_{k} \beta_{k} \ket{0},
\end{equation}
where $\ket{0}$ is the particle vacuum and $\beta_{k}$ are the quasiparticle
annihilation operators as determined by the Bogoliubov transformation. In
practice, the HFB equation with blocking is solved by substituting the column
vectors $(U_\alpha, V_\alpha) \leftrightarrow (V_\alpha^*, U_\alpha^*)$ for the
quasiparticle $\alpha$ one wishes to block \cite{mang1975selfconsistent,
schunck2019energy}. This procedure is performed at each iteration of the
self-consistent loop. The density matrix and pairing tensor in configuration
space become
\begin{subequations}
\begin{align}
\rho^{B,\alpha}_{ij}   & = \big( V^{*}V^{T} \big)_{ij} + U_{i\alpha}U_{j\alpha}^{*} - V_{i\alpha}^{*}V_{j\alpha}, \\
\kappa^{B,\alpha}_{ij} & = \big( V^{*}U^{T} \big)_{ij} + U_{i\alpha}V_{j\alpha}^{*} - V_{i\alpha}^{*}U_{j\alpha}.
\end{align}
\end{subequations}

The exact implementation of the blocking prescription breaks time-reversal
symmetry and depends on the self-consistent symmetries
\cite{schunck2010onequasiparticle}. For this reason, one often employs the
equal filling approximation (EFA) where time-reversal symmetry is explicitly
enforced \cite{perez-martin2008microscopic}. Detailed comparisons of the
energies of blocking configurations near the Fermi level showed that the error
incurred when using the EFA does not exceed a few keV \cite{duguet2001pairing,
schunck2010onequasiparticle}. As demonstrated in
Ref.~\cite{perez-martin2008microscopic}, the EFA can be thought of as a special
statistical mixture of one-quasiparticle states. The density matrices are thus 
modified to read as
\begin{subequations}
\begin{align}
\rho^{{\rm EFA},\alpha}_{ij}
= \big( V^{*}V^{T} \big)_{ij} + \frac{1}{2}\big( 
  & U_{i\alpha}U_{j\alpha}^{*} - V_{i\alpha}^{*}V_{j\alpha} \nonumber\\
+ ~ & U_{i\bar{\alpha}}U_{j\bar{\alpha}}^{*} - V_{i\bar{\alpha}}^{*}V_{j\bar{\alpha}} \big),\\
\kappa^{{\rm EFA},\alpha}_{ij}
= \big( V^{*}U^{T} \big)_{ij} + \frac{1}{2}\big( 
  & U_{i\alpha}V_{j\alpha}^{*} - V_{i\alpha}^{*}U_{j\alpha} \nonumber\\
+ ~ & U_{i\bar{\alpha}}V_{j\bar{\alpha}}^{*} - V_{i\bar{\alpha}}^{*}U_{j\bar{\alpha}} \big).
\end{align}
\end{subequations}

In this work, the selection of quasiparticle states to block follows the
automated procedure outlined in Ref.~\cite{stoitsov2013axially}. From the HFB
solution in the even-even neighbor, the code identifies an initial set
$\mathcal{B}$ of blocking candidates $\alpha$ within a small energy window
around the Fermi energy by imposing the condition $|E_{\alpha} - E_{0}| \leq
\Delta E$, where $E_0$ is the energy of the lowest quasiparticle and $\Delta E$ 
is the energy window. This procedure is applied for each $\Omega$-block. At 
each iteration $n$, the code computes the overlap $O_{\alpha\alpha'}$ between 
the blocked state $\alpha \equiv \alpha^{(n-1)}$ at the previous iteration and 
each quasiparticle state in the same $\Omega$-block, $O_{\alpha\alpha'} = 
\sum_{i} \big( U_{i\alpha}U_{i\alpha'} + V_{i\alpha}V_{i\alpha'} \big)$. The 
quasiparticle $\alpha'$ with the maximum overlap defines the updated version of 
$\alpha$ at iteration $n$, $\alpha^{(n)} = \alpha'$.

Applying this blocking prescription at each point $\qvec$ of the PES gives a
set of $N_{q}$ blocking potential energies, $\mathcal{S}_q \equiv
\{ V_{\alpha}(\qvec) \}_{\alpha=1,\dots,N_q}$. Note that the number $N_q$ of 
such configurations is not the same everywhere on the PES, since the blocking
criterion is based on energy. Similarly, the numbers of blocked states 
$N_{\Omega}$ with $\Omega = 1/2$, $\Omega = 3/2$, $\dots$, within a given set
$\mathcal{S}_q$ are not identical. Typically the number of blocked states is
such that $N_{1/2} \geq N_{3/2} \geq\dots$.

\subsubsection{Time-Dependent Generator Coordinate Method}
\label{subsubsec:tdgm}

In this work, we assume that the large-amplitude collective dynamics of the 
fissioning nucleus can be approximated by the time-dependent generator 
coordinate method (TDGCM) under the Gaussian overlap approximation (GOA). Let 
us recall that the fundamental hypothesis of this method is that the quantum 
state $\ket{{\rm GCM}(t)}$ that describes the fissioning system is a 
time-dependent linear superposition of static states at different deformations
\begin{equation}
  \ket{{\rm GCM}(t)}=\int d\qvec \ket{\Phi(\qvec)} f(\qvec, t),
\end{equation}
where $f$ is a complex-valued function that defines the superposition at each 
time $t$ and $\ket{\Phi(\qvec)}$ is a constrained HFB state. Using the 
additional hypotheses of the GOA, the TDGCM+GOA equation of motion reads
\begin{equation}
\label{eq:tdgcmgoa}
  i\hbar \pdv{\GCMstate(\qvec,t)}{t}
  = \left[
      \Hcoll(\qvec)
    + i\Acoll(\qvec)
    \right] \, \GCMstate(\qvec,t),
\end{equation}
where the complex-valued function $\GCMstate(\qvec,t)$, equivalent to 
$f(\qvec, t)$, contains all the information about the dynamics of the system 
and $\Acoll(\qvec)$ is a real-valued field that is added to avoid reflection on 
the boundaries of the deformation domain~\cite{regnier2018felix2}. The 
collective Hamiltonian $\Hcoll(\qvec)$ is a local linear operator acting on
$\GCMstate(\qvec,t)$,
\begin{equation}
\label{eq:Hcoll}
\Hcoll(\qvec) \equiv
-\frac{\hbar ^2}{2\gamma^{1/2}(\qvec)} 
\sum_{\mu\nu} \pdv{q_{\mu}} \gamma^{1/2}(\qvec) B_{\mu\nu}(\qvec) \pdv{q_{\nu}} 
+  V(\qvec),
\end{equation}
with $B_{\mu\nu}(\qvec)$ the components of the collective inertia tensor, 
$V(\qvec)$ the potential energy, which is the sum of the HFB energy and some
zero-point energy corrections, and $\gamma$ is the GCM metric 
\cite{krappe2012theory}.

Equation \eqref{eq:tdgcmgoa} is derived from the Hill-Wheeler-Griffin equation
of the GCM after applying the GOA \cite{verriere2020timedependent}. The
derivation does require that the generator states $\ket{\Phi(\qvec)}$ are pure 
states with the structure of quasiparticle vacuum \cite{schunck2019energy}. In 
the EFA, this is not satisfied since the system is in fact described by a (very
specific) statistical density operator. However, it is possible to compute
every ingredient of Eq.\eqref{eq:Hcoll} (potential energy, zero-point energy
corrections, and collective inertia) for statistical ensembles through the
extension of the adiabatic time-dependent Hartree-Fock-Bogoliubov (ATDHFB) at
finite temperature. Therefore, we adopt the pragmatic point of view of using 
Eq.~\eqref{eq:tdgcmgoa} as the equation of motion with inputs determined from 
the finite-temperature ATDHFB theory -- with statistical occupations given by 
the EFA prescription. Since the (TD)GCM formalism has not been extended to 
finite temperature yet, this is a reasonable compromise that has already been 
applied to study the structure of the collective inertia mass tensor as a
function of temperature \cite{martin2009fission}, thermal spontaneous fission
rates \cite{zhu2016thermal}, the dependency of primary fission fragment
distributions on excitation energy \cite{zhao2019microscopic}, and to estimate
dissipation effects in fission fragment distributions 
\cite{zhao2022timedependent}. In spite of these examples, the full derivation
of the ATDHFB collective inertia tensor $B_{ij}(\qvec)$ at finite temperature 
has never been presented. The special case of the ATDHFB+BCS inertia was 
derived in Refs.~\cite{iwamoto1979behavior,baran1994temperature} by replacing 
expectation values of observables in the zero-temperature cranking model 
formula by ensemble averages and the full, correct ATDHFB result was given 
without proof in Ref.~\cite{martin2009fission}. Therefore, we demonstrate below 
how to obtain the formula for the finite-temperature ATDHFB collective inertia 
tensor.

\paragraph{ATDHFB equations}
The starting point of the derivation is the Liouville equation for the density
operator $\mathcal{D}$ \cite{sommermann1983microscopic}. Applying the
statistical Wick theorem yields the finite-temperature time-dependent
Hartree-Fock-Bogoliubov (TDHFB) equation, which is formally equivalent to the
zero-temperature TDHFB equation, $i\hbar\dot{\mathcal{R}} =
[ \mathcal{H}, \mathcal{R}]$ \cite{blaizot1985quantum}. Following the ideas of
ATDHFB \cite{baranger1978adiabatic}, we then write the TDHFB generalized
density $\mathcal{R}(t)= e^{i\chi(t)} \mathcal{R}^{(0)}(t)e^{-i\chi(t)}$, where
$\chi(t)$ is a quadratic form of single-particle creation and annihilation
operators \cite{blaizot1985quantum}. The Taylor series expansion of $\chi(t)$
up to second order, $\chi(t) = \chi^{(0)}(t) + \chi^{(1)}(t) + \chi^{(2)}(t)$,
induces a similar expansion for the TDHFB generalized density matrix,
$\mathcal{R}(t)= \mathcal{R}^{(0)}(t) + \mathcal{R}^{(1)}(t) +
\mathcal{R}^{(2)}(t)$, for example: $\mathcal{R}^{(1)} =
i\big[ \chi(t), \mathcal{R}^{(0)}(t) \big]$. Plugging these two Taylor
expansions into the finite-temperature TDHFB equation and separating 
contributions that are time-even from the ones that are time-odd gives a set of 
coupled equations that are formally analogous to the zero-temperature ATDHFB 
equations,
\begin{subequations}
\begin{align}
\displaystyle i\hbar\dot{\mathcal{R}}^{(0)}
& = [ \mathcal{H}^{(0)}, \mathcal{R}^{(1)} ] + [ \mathcal{H}^{(1)}, \mathcal{R}^{(0)} ], \label{eq:ATDHFB_0}
\medskip\\
\displaystyle i\hbar\dot{\mathcal{R}}^{(1)}
& = [ \mathcal{H}^{(0)}, \mathcal{R}^{(0)} ] + [ \mathcal{H}^{(0)}, \mathcal{R}^{(2)} ] \nonumber\\
& + [ \mathcal{H}^{(1)}, \mathcal{R}^{(1)} ] + [ \mathcal{H}^{(2)}, \mathcal{R}^{(0)} ] . \label{eq:ATDHFB_1}
\end{align}
\end{subequations}
In Eqs.~(\ref{eq:ATDHFB_0}-\ref{eq:ATDHFB_1}), the matrices $\mathcal{H}^{(n)}$
represent the FT-HFB matrices at order $n$, i.e., they depend on the order-$n$
density matrices $\rho^{(n)}(t)$ and $\kappa^{(n)}(t)$ that enter the
generalized densities $\mathcal{R}^{(n)}(t)$,
\begin{align}
\mathcal{R}^{(0)}(t) & =
\left( \begin{array}{cc}
\rho^{(0)} & \kappa^{(0)} \\ -\kappa^{(0)*} & 1-\rho^{(0)*}
\end{array}\right),
\medskip\\
\mathcal{R}^{(n)}(t) & =
\left( \begin{array}{cc}
\rho^{(n)} & \kappa^{(n)} \\ -\kappa^{(n)*} & -\rho^{(n)*}
\end{array}\right).
\end{align}
We now introduce the TDHFB quasiparticle basis, which diagonalizes at each time 
the  zero-order, finite-temperature $\mathcal{R}^{(0)}(t)$ density matrix. In 
that basis, $\mathcal{R}^{(0)}(t)$ takes the form
\begin{equation}
\tilde{\mathcal{R}}^{(0)}(t)
= \left( \begin{array}{cc} f(t) & 0 \\ 0 & 1-f(t) \end{array}\right),
\end{equation}
with $f_{kl}(t)$ the statistical occupation factors. In the following, the
tilde indicates that a matrix is written in the TDHFB quasiparticle basis.

\paragraph{Energy to Second Order}
The next step is to obtain a self-contained expression for the total energy of
the system that only depends on the operator $\chi(t)$. This is achieved by
first expanding the TDHFB energy in terms of the matrices $\mathcal{R}^{(n)}$.
One obtains
\begin{equation}
E = E_{\rm HFB}
+ \frac{1}{2} \trace( \mathcal{H}^{(0)}\mathcal{R}^{(2)} )
+ \frac{1}{4} \trace( \mathcal{H}^{(1)}\mathcal{R}^{(1)} ) .
\end{equation}
We can then use the commutators that relate each of the $\mathcal{R}^{(n)}$ to
$\chi(t)$ to obtain an expression of $E$ as a function of $\chi(t)$ only. Even
when working in the TDHFB quasiparticle basis, the full calculation is rather 
lengthy because of the term proportional to $\mathcal{H}^{(1)}$, which depends 
on $\mathcal{R}^{(1)}$ indirectly through its components $\rho^{(1)}$ and 
$\kappa^{(1)}$ and corresponds to the off-diagonal terms of the FT-QRPA matrix. 
Since we work at the cranking approximation, we neglect it. It is then 
relatively straightforward to show that the total energy reduces to
\begin{equation}
E = E_{\rm HFB} + \frac{1}{4} \vec{\chi}^{\dagger} \mathcal{M} \vec{\chi},
\end{equation}
where $\vec{\chi}$ is the linearized version of the matrix of the operator
$\chi$ in the TDHFB quasiparticle basis,
\begin{equation}
\tilde{\chi} = \left(\begin{array}{cc}
\tilde{\chi}^{11} & \tilde{\chi}^{12} \\
\tilde{\chi}^{21} & \tilde{\chi}^{22}
\end{array}\right)
\quad\Rightarrow \quad
\vec{\chi} =
\left( \begin{array}{c}
\tilde{\chi}^{11} \\ \tilde{\chi}^{12} \\ \tilde{\chi}^{21} \\ \tilde{\chi}^{22}
\end{array}\right),
\end{equation}
and $\mathcal{M}$ is the FT-QRPA matrix in the cranking approximation. In that 
same linearized TDHFB basis, it can be written 
$\mathcal{M} = \mathbb{E}\mathbb{F}$ with
\begin{equation*}
\setlength{\arraycolsep}{2.5pt}
\medmuskip =1mu
\mathbb{E} =
\left( \begin{array}{cccc}
 (E_k - E_l) & & & \\
 & (E_k + E_l) & &  \\
 & & -(E_k + E_l) & \\
 & & & - (E_k - E_l)
\end{array}\right)
\end{equation*}
and
\begin{equation*}
\setlength{\arraycolsep}{2.5pt}
\medmuskip =1mu
\mathbb{F} =
\left( \begin{array}{cccc}
-(f_k - f_l) & & & \\
 & (1 - f_k - f_l) & &  \\
 & & -(1 - f_k - f_l) & \\
 & & & (f_k - f_l)
\end{array}\right).
\end{equation*}
In these last two expressions, terms like $E_k - E_l$ stand for the matrix 
$\tilde{E}$ with elements $\tilde{E}_{kl} = E_k -E_l$ with $E_{k}$ the 
quasiparticle energies.

\paragraph{Adiabatic Approximation}
Starting from Eqs.(\ref{eq:ATDHFB_0}\ref{eq:ATDHFB_1}) and continuing to work 
in the TDHFB quasiparticle basis, one can show that
\begin{equation}
\hbar\vec{\dot{\mathcal{R}}}^{(0)}
= \mathbb{E}\mathbb{F}\vec{\chi}
= \mathcal{M}\vec{\chi} ,
\end{equation}
where $\vec{\dot{\mathcal{R}}}^{(0)}$ is, again, the linearized matrix of the
operator $\dot{\mathcal{R}}^{(0)}$ in the TDHFB basis. The total energy thus
reads
\begin{equation}
E = E_{\rm HFB} + \frac{\hbar^2}{4} \vec{\dot{\mathcal{R}}}^{(0)\dagger} \mathcal{M}^{-1} \vec{\dot{\mathcal{R}}}^{(0)}.
\end{equation}

As is customary in practical applications of the ATDHF or ATDHFB theory
\cite{baranger1978adiabatic}, we then introduce a (small) set of collective
variables $\qvec \equiv (q_1, \dots, q_{N})$ and assume that the densities
$\mathcal{R}^{(0)}(t)$ vary in time only through changes in these collective
variables,
\begin{equation}
\dot{\mathcal{R}}^{(0)} = \sum_{\mu} \dot{q}_{\mu} \pdv{\mathcal{R}^{(0)}}{q_{\mu}} .
\end{equation}
In physics terms, this statement is the equivalent of the Born-Oppenheimer
approximation: the nuclear dynamics is confined to a collective space. 
Additionally, we approximate the solutions of the finite-temperature HFB
equation constrained on the expectation value $\qvec$ of the collective
variables by the static densities $\mathcal{R}^{(0)}$. In other words, the 
collective space that contains the nuclear dynamics is precalculated as a 
series of FT-HFB calculations. Let us emphasize here that these approximations 
are exactly the same as the ones underpinning the zero-temperature expressions 
of the ATDHFB collective inertia that are commonly used in the literature.

\paragraph{Local Approximation}
The final stage of the derivation consists in expressing
$\pdv*{\mathcal{R}^{(0)}}{q_{\mu}}$ locally at point $\qvec$. Since we have 
assumed that the density $\mathcal{R}^{(0)}$ is the solution of the FT-HFB 
equation with constraints $\qvec$, it satisfies 
$\big[ \mathcal{H}^{(0)} - \sum_{\mu} \lambda_{\mu} \ope{Q}_{\mu},
\mathcal{R}^{(0)} \big] = 0$ with $\lambda_{\mu}$ the Lagrange parameter
associated with the constraint operator $\ope{Q}_{\mu}$. We collect all such
parameters into the vector $\gras{\lambda} = (\lambda_1,\dots, \lambda_{N})$.
We then express that this equation must be satisfied for small variations of
the density, that is, when
\begin{align*}
\mathcal{R}^{(0)} &\rightarrow \mathcal{R}^{(0)} + \mathcal{R}^{(1)}, \\
\mathcal{H}^{(0)} &\rightarrow \mathcal{H}^{(0)} + \mathcal{H}^{(1)}, \\
\lambda_{\mu} &\rightarrow \lambda_{\mu} + \delta\lambda_{\mu}.
\end{align*}
Introducing these expansions into the FT-HFB equation with constraints and
taking advantage of the quasiparticle basis, some simple algebra leads to the
following relation: $\vec{\mathcal{R}}^{(1)} = -\delta\gras{\lambda}\cdot
\mathbb{E}^{-1} \mathbb{F} \vec{\vb{Q}}$ where $\vec{\vb{Q}} \equiv 
(\vec{Q}_1,\dots,\vec{Q}_{N})$ is a vector containing the linearized matrix 
$\vec{Q}_{\mu}$ of the constraint operator in the TDHFB quasiparticle basis. To 
clarify, the condensed notation stands for
\begin{align*}
\vec{\mathcal{R}}^{(1)}
& = -\delta\gras{\lambda}\cdot \mathbb{E}^{-1} \mathbb{F} \vec{\vb{Q}} \\
& = -\sum_{\mu} \delta \lambda_{\mu} \sum_{kl}
\Big[
 \frac{f_k-f_l}{E_k-E_l}  \tilde{Q}_{\mu;kl}^{11}
-\frac{1-f_k-f_l}{E_k+E_l}\tilde{Q}_{\mu;kl}^{12} \\
&
-\frac{1-f_k-f_l}{E_k+E_l}\tilde{Q}_{\mu;kl}^{21}
+\frac{f_k-f_l}{E_k-E_l}  \tilde{Q}_{\mu;kl}^{22}
\Big].
\end{align*}
We apply the chain rule to write
\begin{equation}
\pdv{\mathcal{R}^{(0)}}{q_{\mu}}
= \sum_{\alpha}
\fdv{\mathcal{R}^{(0)}}{\lambda_{\alpha}}
\fdv{\lambda_{\alpha}}{q_{\mu}}.
\end{equation}
At this point, we identify the small variations of the generalized density with
the first-order variations $\mathcal{R}^{(1)}$, i.e., $\delta\mathcal{R}^{(0)}
= \mathcal{R}^{(1)}$. We then obtain the variations $\delta\lambda_a$ simply
by recalling that in the TDHFB quasiparticle basis,
\begin{equation}
  q_{\mu} =
    \frac{1}{2}\trace( Q_{\mu})
  + \frac{1}{2}\trace( Q_{\mu}\mathcal{R}^{(0)})
  \Rightarrow
  \delta q_{\mu} =
  \frac{1}{2} \trace( Q_{\mu}\mathcal{R}^{(1)}).
\end{equation}
The variation can also be written $\delta q_{\mu} = \frac{1}{2}
\vec{Q}_{\mu}^{\dagger}\vec{\mathcal{R}}^{(1)}$. Using the previous 
relationship between $\vec{\mathcal{R}}^{(1)}$ and $\vec{\vb{Q}}$, we can
find that
\begin{equation}
\delta\lambda_{\nu} = 2\sum_{\alpha} \big[ \mathsf{M} \big]^{-1}_{\nu\alpha} \delta q_{\alpha}
\end{equation}
with the moments $\mathsf{M}^{(K)} \equiv\vec{\vb{Q}}^{\dagger}
\mathbb{E}^{-K}\mathbb{F}\vec{\vb{Q}}$, that is,
\begin{align*}
\mathsf{M}_{\mu\nu}^{(K)}
& = \sum_{kl}
\Big[
 \tilde{Q}_{\mu;kl}^{11*}\frac{f_l-f_k}{(E_k-E_l)^{K}}  \tilde{Q}_{\nu;kl}^{11}
+\tilde{Q}_{\mu;kl}^{12*}\frac{1-f_l-f_k}{(E_k+E_l)^{K}}\tilde{Q}_{\nu;kl}^{12}
\phantom{\Big].}
\medskip\\
&
+\tilde{Q}_{\mu;kl}^{21*}\frac{1-f_l-f_k}{(E_k+E_l)^{K}}\tilde{Q}_{\nu;kl}^{21}
+\tilde{Q}_{\mu;kl}^{22*}\frac{f_l-f_k}{(E_k-E_l)^{K}}  \tilde{Q}_{\nu;kl}^{22}
\Big].
\end{align*}
The time variations of $\mathcal{R}^{(0)}$ thus become
\begin{equation}
\dot{\vec{\mathcal{R}}}^{(0)}
=
2\sum_{\alpha\beta} \dot{q}_{\beta}\big[\mathsf{M}^{(1)}\big]^{-1}_{\alpha\beta}
\mathbb{E}^{-1}\mathbb{F}\vec{Q}_{\beta}
\end{equation}
leading to the total energy taking the form
\begin{equation}
E_{2} = \frac{1}{2} \sum_{\mu\nu} \mathsf{M}_{\mu\nu} \dot{q}_{\mu}\dot{q}_{\nu}
\end{equation}
with
\begin{equation}
\mathsf{M} = 2\hbar^{2} \big[\mathsf{M}^{(1)}\big]^{-1}\mathsf{M}^{(3)}\big[\mathsf{M}^{(1)}\big]^{-1}.
\label{eq:inertia_tensor}
\end{equation}
Apart from a factor of 2, this formula is the same as the zero-temperature 
result. The main difference lies in the definition of the moments 
$\mathsf{M}_{\mu\nu}^{(K)}$, which depend explicitly on the Fermi-Dirac
statistical occupations. In the case of the EFA, we recall that $f_k = 0$ 
except $f_{\alpha}= f_{\bar{\alpha}}= 1/2$.

\section{Static Potential Energy Surfaces}
\label{sec:static}

As mentioned earlier, we assume in this work that axial and time-reversal 
symmetries are conserved. In addition to accelerating calculations 
substantially, this hypothesis greatly facilitates the implementation of the 
blocking prescription as discussed in Section \ref{subsubsec:blocking}. 
Enforcing axial symmetry has two main consequences: (i) K-mixing between states 
is not possible and (ii) the height of the first fission barrier will be 
systematically overestimated by about 1-1.5 MeV \cite{larsson1972fission,
burvenich2004systematics,kowal2010fission,abusara2010fission,lu2012potential,
abusara2012fission,schunck2014description}. 

All calculations were performed with the code HFBTHO 
\cite{marevic2022axiallydeformed}. We use a deformed harmonic oscillator (HO) 
basis containing up to $N_{\rm shells} = 30$ and truncated to $N_{\rm states} 
= 1100$. The HO spherical frequency $\omega_0$ and its axial deformation 
$\beta_2$ were adjusted based on the value of the quadrupole moment $\quadru$ 
following the empirical formula presented in \cite{schunck2014description}. We 
used the SkM* parameterization of the Skyrme functional \cite{bartel1982better} 
and a surface-volume, zero-range, density-dependent pairing interaction with a 
cut-off $E_{\rm cut} = 60$ MeV. The neutron and proton strengths of the pairing 
force were adjusted to the three-point odd-even staggering in $^{236}$U: 
$V_{n} = -255.250$ MeV and $V_{p} = -325.594$ MeV.

\subsection{One-Dimensional Fission Paths}
\label{subsec:1d}

We begin by recalling the role of quasiparticle blocking on one-dimensional
fission paths. Calculations of fission barriers in odd-mass nuclei were first
reported within the microscopic-macroscopic model using the blocking 
prescription \cite{nilsson1969new,randrup1973theoretical,cwiok1985fission,
jachimowicz2017adiabatic}. Fully self-consistent calculations of fission paths 
in odd nuclei were performed for the Gogny force \cite{perez-martin2009fission,
rodriguez-guzman2017microscopic}. Most of these calculations focused on which 
$K$-value gives the lowest fission barrier or lowest energy fission path. In 
Refs.~\cite{nilsson1969new,randrup1973theoretical}, the authors investigated 
how changes in the blocking configuration affected the height of the barrier in 
a few select cases. Similar calculations were performed in the Hartree-Fock 
plus BCS formalism in Refs.~\cite{hock2013fission,koh2017fission}, where the 
authors mentioned the consequences of the variations in fission barriers on 
quantities such as fission penetrabilities, which enter fission cross-section 
models. 

\begin{figure}[!ht]
\begin{center}
\includegraphics[width=\columnwidth]{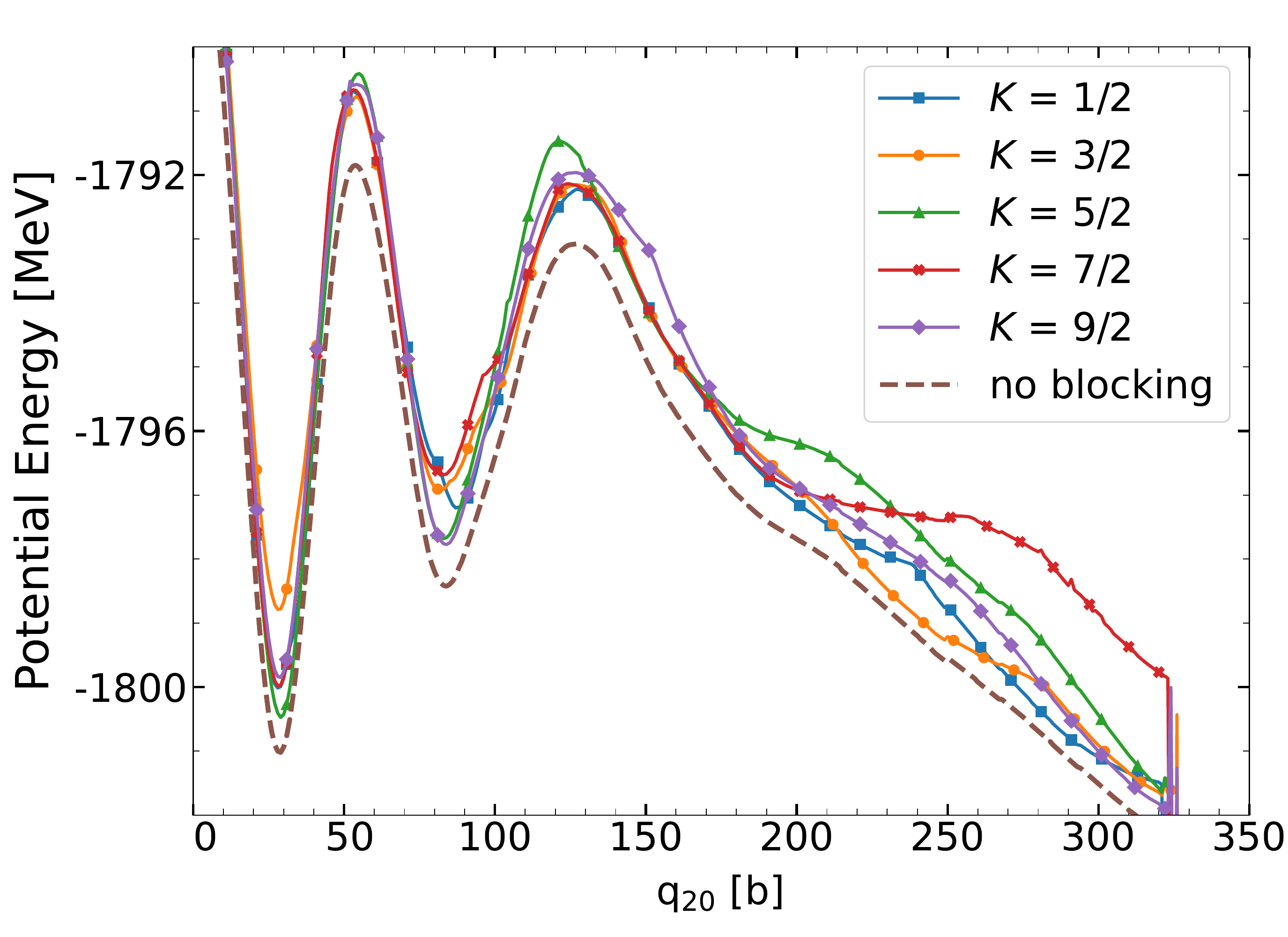}
\caption{\label{fig:u239_blocking} Potential energy curves in $^{239}$U as a
function of the axial quadrupole moment $\quadru$ for different blocking
configurations $K=1/2,\dots,9/2$. The dashed line corresponds to a HFB
calculation without the blocking prescription where the average number of
particles is set to $Z_0 = 92$ and $N_0 = 147$.}
\end{center}
\end{figure}

In this section, we perform a more systematic exploration of the dependency of 
the full potential energy curves, from the ground state to the scission point, 
for different values of $K$. Figure \ref{fig:u239_blocking} shows the example 
of blocking calculations in $^{239}$U. For all $K$-values included in the 
figure, the curve shows the energy of the lowest blocked configuration having 
that given $K$ as a function of the expectation value of the axial quadrupole 
moment $\quadru$. For comparison, we also show the result obtained in the 
no-blocking approximation, where the HFB equation was solved by simply 
constraining the average value of $N$ to be $ \expval*{\hat{N}} = 147$.
We first note that the spin of the ground state (g.s.) is $K=5/2$, which agrees 
with experimental assignment \cite{nndc}. For $^{237}$U, we found $K=1/2$ for 
the g.s., which is also in agreement with experimental results. As already 
noticed in Refs.~\cite{perez-martin2009fission,
rodriguez-guzman2017microscopic}, the potential energy of blocked 
configurations is systematically higher than the no-blocking ones, which is a
manifestation of the `specialization' effect \cite{koh2017fission}. 

The comparison of the fission paths for different values of $K$ show 
significant differences of the order of up to several MeV. Table 
\ref{tab:barriers} lists the height of the first fission barrier ($\EA$) and 
second fission barrier ($\EB$) as well as the excitation energy of the fission
isomer ($\EFI$) for both $^{237}$U and $^{239}$U. The maximum difference 
reaches 1.28 MeV for $\EA$, 1.45 MeV for $\EB$ and 1.90 MeV for $\EFI$ for 
$^{237}$U, and 2.05 MeV for $\EA$, 2.39 MeV for $\EB$ and 1.54 MeV for $\EFI$ 
for $^{239}$U. Interestingly, and perhaps coincidentally, for both nuclei the 
mean value of $\EA$, $\EB$ and $\EFI$ over the range of $K$ values is quite 
close to the no-blocking result. These results confirm the conclusions in 
Ref.~\cite{koh2017fission}: since fission barrier heights enter in the form of 
an exponential in standard formulas for either spontaneous fission half-lives 
or fission cross sections \cite{younes2021introduction,schunck2022theory}, such 
differences are actually considerable. 

\begin{table}[!htb]
\caption{\label{tab:barriers} Characteristic points of potential energy curves
in $^{237}$U and $^{239}$U: height of the first barrier ($\EA$), of the second
barrier ($\EB$), and excitation energy of the fission isomer ($\EFI$). The
first five columns correspond to blocking configurations characterized by the
$K$ quantum number; the last column stands for the no-blocking option. All
values are given in MeV.}
\begin{ruledtabular}
\begin{tabular}{llcccccc}
         &  $K$   &  1/2 &  3/2 &  5/2  &  7/2 &  9/2 & no blck.\\
\hline
$^{237}$U          & $\EA$  & 9.17 & 7.89 &  9.09 & 9.00 & 8.17 & 8.54 \\
          & $\EB$  & 7.77 & 6.58 &  8.03 & 7.68 & 7.05 & 7.32 \\
          & $\EFI$ & 3.63 & 2.83 &  3.56 & 4.62 & 2.72 & 3.45 \\
\hline
$^{239}$U & $\EA$  & 9.31 & 8.00 & 10.05 & 9.30 & 9.27 & 9.17 \\
          & $\EB$  & 7.78 & 6.63 &  8.99 & 7.83 & 7.88 & 7.94 \\
          & $\EFI$ & 2.86 & 2.34 &  3.49 & 3.88 & 2.67 & 3.09 \\
\end{tabular}
\end{ruledtabular}
\end{table}

There were many studies of the evolution of fission barriers with angular 
momentum \cite{mustafa1974angular,faessler1981description,diebel1981microscopic,
nemeth1985hartreefock,garcias1990angular,egido2000fission,baran2014fission}. 
Irrespective of the details of the theoretical model employed, all results 
pointed to the gradual decrease of the barriers due to the damping of shell 
effects. However, these analyses were focused on the total angular momentum $J$ 
of even-even nuclei in a rather high-spin regime. Our axially-symmetric 
blocking calculations only provide the eigenvalue $K$ of $\hat{J}_z$ and we 
have $J \geq K$. Even though the $K$ dependency of fission barriers in odd-mass 
nuclei as captured by blocking calculations is nonlinear -- the height of the 
barrier is maximum at $K=1/2$ for $^{237}$U but at $K=5/2$ for $^{239}$U -- one 
cannot exclude that full angular momentum projection would restore the order 
that one might expect from even-even nuclei ($\EA(J=\tfrac{1}{2}) 
\geq \EA(J=\tfrac{3}{2}) \geq ...$).

While relative differences, as quantified by the height of fission barriers,
are large, absolute differences are much smaller: the energy at the top of the
first barrier does not vary by more than 180 keV in $^{237}$U and 280 keV in
$^{239}$U. In contrast, Fig.~\ref{fig:u239_blocking} shows that the energy in
the ground-state potential well at $\quadru \approx 30$ b, or in the descent
from saddle to scission for $\quadru > 170$ b, varies by up to several MeV. If 
such a pattern holds for multi-dimensional potential energy surfaces, these 
results suggest that blocking different $K$-values could have an impact on the 
fission fragment distributions, not just fission probabilities.

Before finishing this section, we should point out a very general limitation
of the blocking prescription in such potential energy surface calculations
(even if it were implemented exactly by breaking time-reversal symmetry and
axial symmetry). As recalled in Section \ref{subsubsec:blocking}, blocking
calculations require a reference state, which is typically chosen as the
neighboring even-even nucleus with either $N-1$ or $N+1$ particles, or 
sometimes the HFB solution for the no-blocking approximation. This prescription 
works very well everywhere except near scission. In one- or two-dimensional
collective spaces, scission often takes the form of a discontinuity in the PES,
as seen at $\quadru \approx 325$ b in Fig.~\ref{fig:u239_blocking}. If this
discontinuity occurs at, say $\quadru^{\rm disc.}= q_0$ for the {\it reference} 
states, then the discontinuity for {\it all} blocking configurations and $K$ 
values built on these reference states must be such that $\quadru^{\rm disc.} 
\leq q_0$. In other words, the blocking scheme cannot produce a PES for some 
$K$ value where scission would occur at larger values of $\quadru$ than in the 
reference state. The only case when such a situation is possible is if the 
collective space is large enough that scission takes place along a continuous 
path.

\subsection{Two-Dimensional Potential Energy Surfaces}
\label{subsec:2d}

\begin{figure}[!ht]
\begin{center}
\includegraphics[width=0.96\columnwidth]{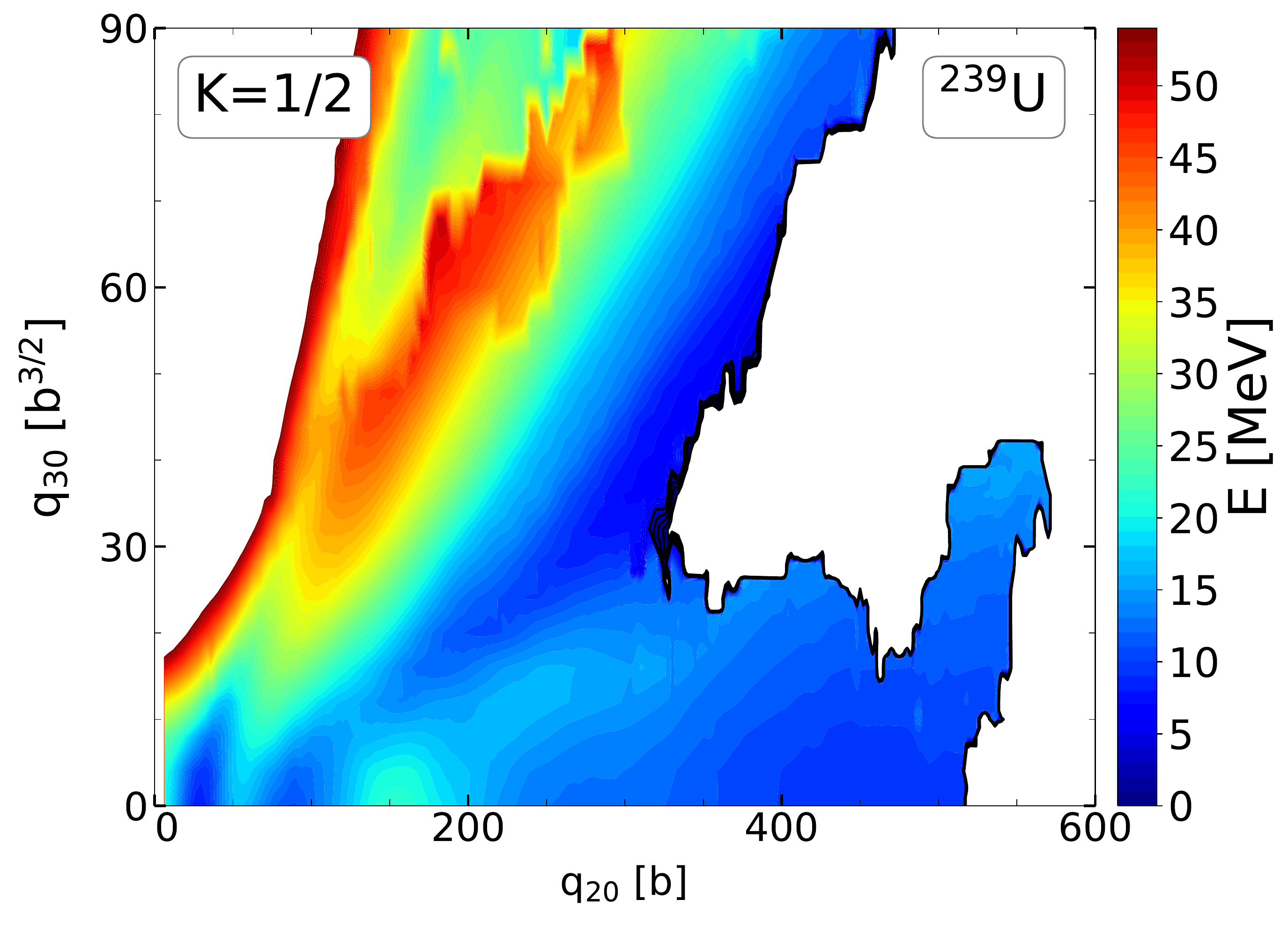}
\caption{Two-dimensional potential energy surface in $^{239}$U as a function of
the expectation value of the axial quadrupole ($\quadru$) and axial octupole
($\octu$) moments for the $K=1/2$ blocking configuration.}
\label{fig:u239_PES_2D}
\end{center}
\end{figure}

The one-dimensional potential energy curves of Sec.~\ref{subsec:1d} can provide
useful meta-data such as barrier heights for the calculation of spontaneous 
fission half-lives or fission cross sections. However, the determination of 
fission fragment distributions requires more collective degrees of freedom. In 
Fig.~\ref{fig:u239_PES_2D}, we show the two-dimensional PES for the $K=1/2$ 
configuration in $^{239}$U. For this nucleus, blocking calculations for all $K$ 
values at the point $\qvec = (\quadru, \octu)$ were initialized from the 
time-even reference state in $^{238}$U at the same point $\qvec$. As mentioned 
in the previous section, this implies that configurations that are beyond 
scission in $^{238}$U are also beyond scission in $^{239}$U. In practice, we 
also found that for nearly all blocking solutions in $^{239}$U, the scission 
line is identical to the one in $^{238}$U. For this reason, the PES for $K=1/2$ 
is, visually, nearly indistinguishable from the ones for $K=3/2,\dots,9/2$ -- 
the color scale would not allow distinguishing differences in energy of the 
order of an MeV -- so we choose to show only one such PES.

This PES is typical of most actinides \cite{dubray2008structure,
younes2009microscopica,schunck2014description,zhao2015multidimensionally,
tao2017microscopic,zhao2019timedependent}: the ground state is
reflection-symmetric and located at around $\quadru\approx 30$ b and the
fission isomer at $\quadru\approx 80$ b (details depend on the nucleus and the
EDF). The second fission barrier is octupole-deformed and leads to the main
fission valley. An additional fission path at much higher energy goes through
very asymmetric shapes associated with cluster radioactivity.

\begin{figure}[!ht]
\begin{center}
\includegraphics[width=0.95\columnwidth]{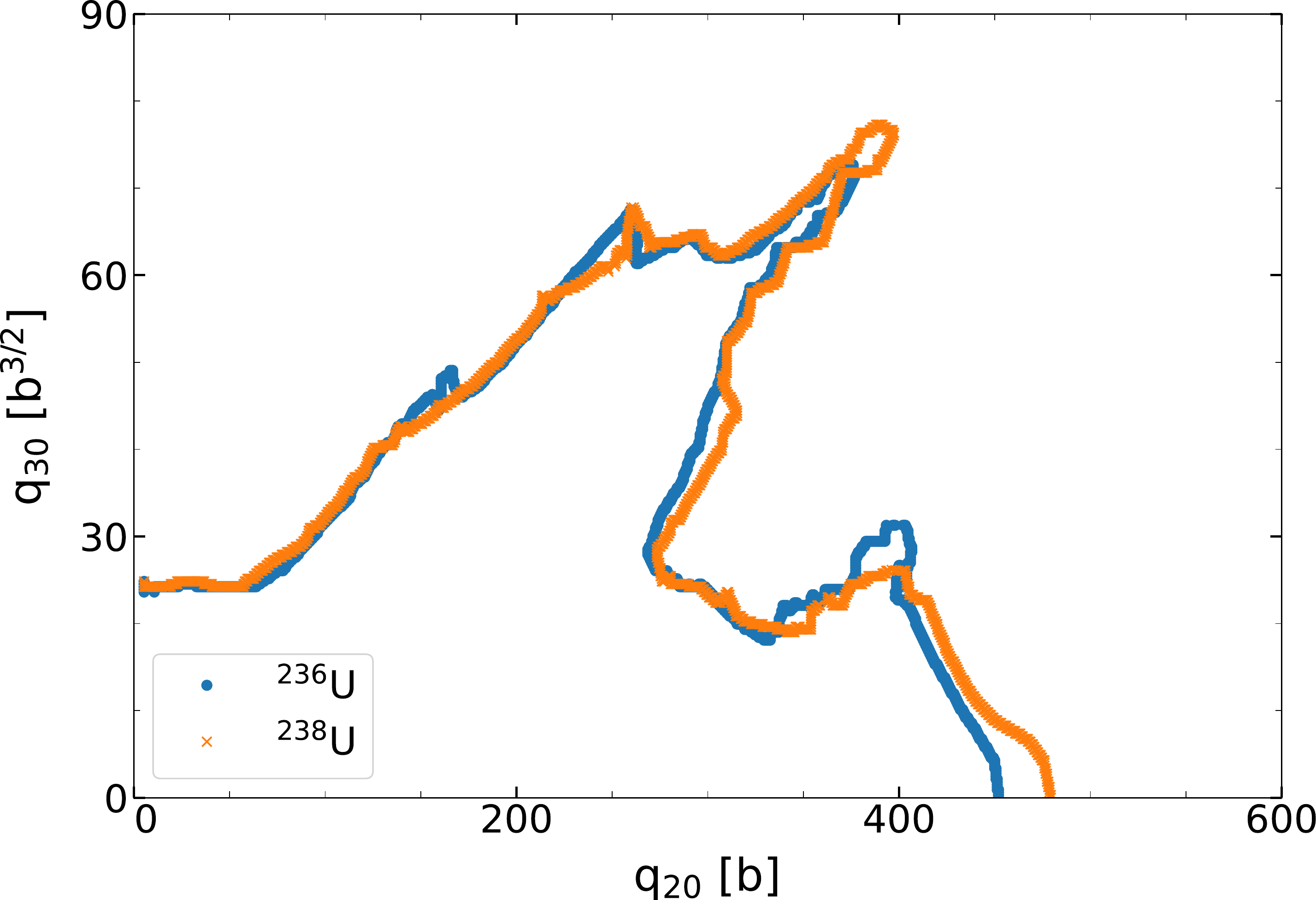}
\caption{Isoline with $\qneck = 6.5$ in the two-dimensional potential energy
surface of $^{236}$U (blue circles) and $^{238}$U (orange crosses).}
\label{fig:u237_scission_line}
\end{center}
\end{figure}

The choice of the time-even reference solutions to initialize blocking
calculations has another consequence. To generate the PES for $^{237}$U, there
are three obvious choices: start from the the PES of $^{236}$U; from the PES of
$^{238}$U; or from the PES of $^{237}$U obtained without blocking. It is
important to realize that the scission line in each of these three PES may be
different. This is illustrated in Fig.~\ref{fig:u237_scission_line}, where the
scission lines of both $^{236}$U and $^{238}$U are represented in the same
figure. In this particular example, we adopted the condition $\qneck = 6.5$ to 
define scission. The most likely fission fragments -- the ones near the peaks 
of the fission fragment distribution -- correspond to the region around 
$\quadru \approx \numrange{300}{350}$ b and $\octu \approx 40$ b$^{3/2}$, i.e., 
to the right-hand side of the figure. Because scission configurations are not 
identical in each nucleus, blocking calculations in $^{237}$U will give 
slightly different results depending on whether the PES for $^{237}$U is 
initialized from the one in $^{236}$U or $^{238}$U.

\begin{figure}[!ht]
\begin{center}
\includegraphics[width=0.98\columnwidth]{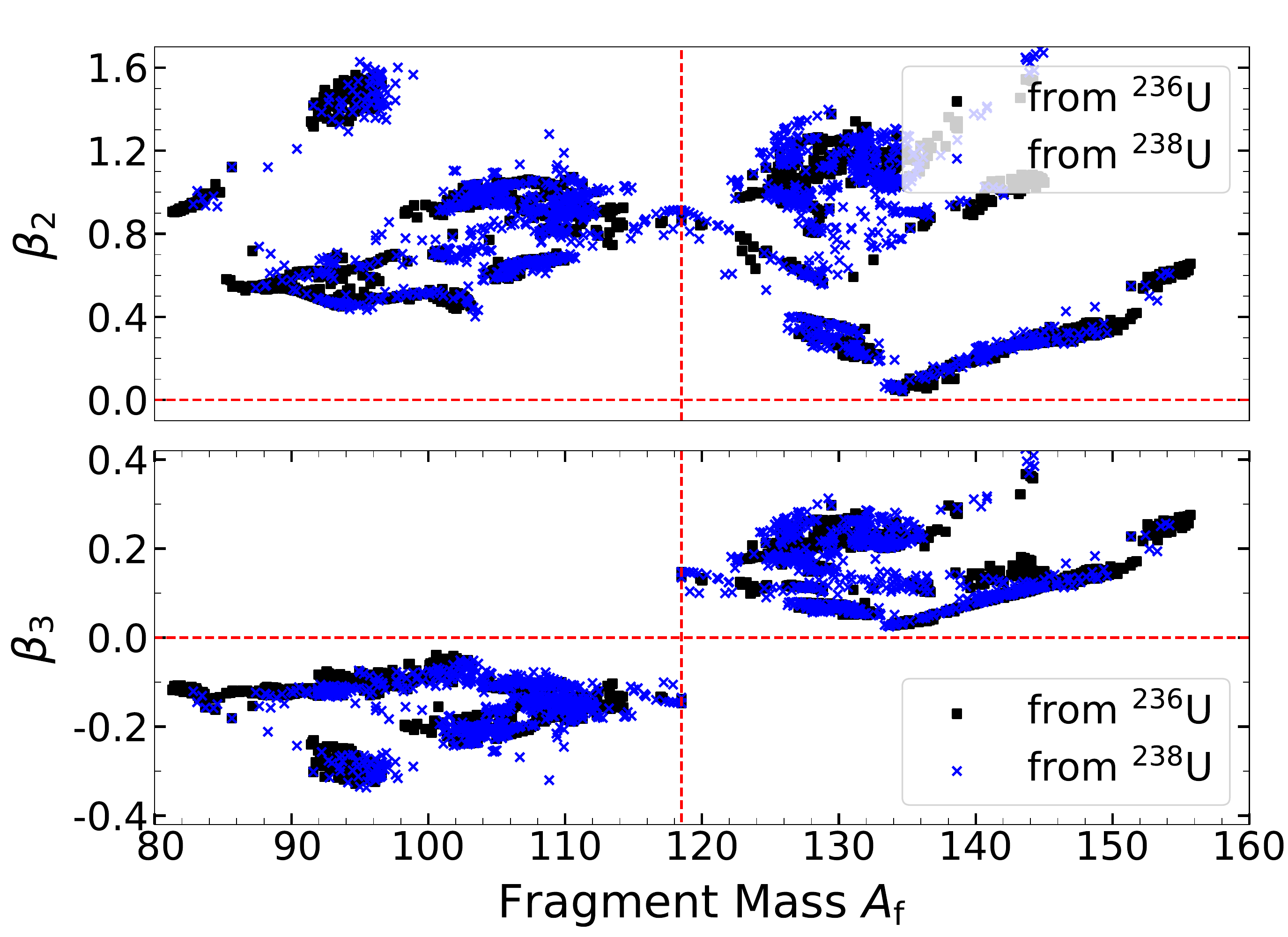}
\caption{Axial quadrupole ($\beta_2$) and octupole ($\beta_3$) deformation in 
the fission fragments in $^{237}$U for $K=1/2$. Black squares represent 
blocking calculations initialized from the PES in $^{236}$U and blue crosses 
represent the ones initialized from the PES in $^{238}$U. The vertical dashed 
line separates the light from the heavy fragments. }
\label{fig:u237_Qlm_frag}
\end{center}
\end{figure}

\begin{figure}[!ht]
\begin{center}
\includegraphics[width=0.98\columnwidth]{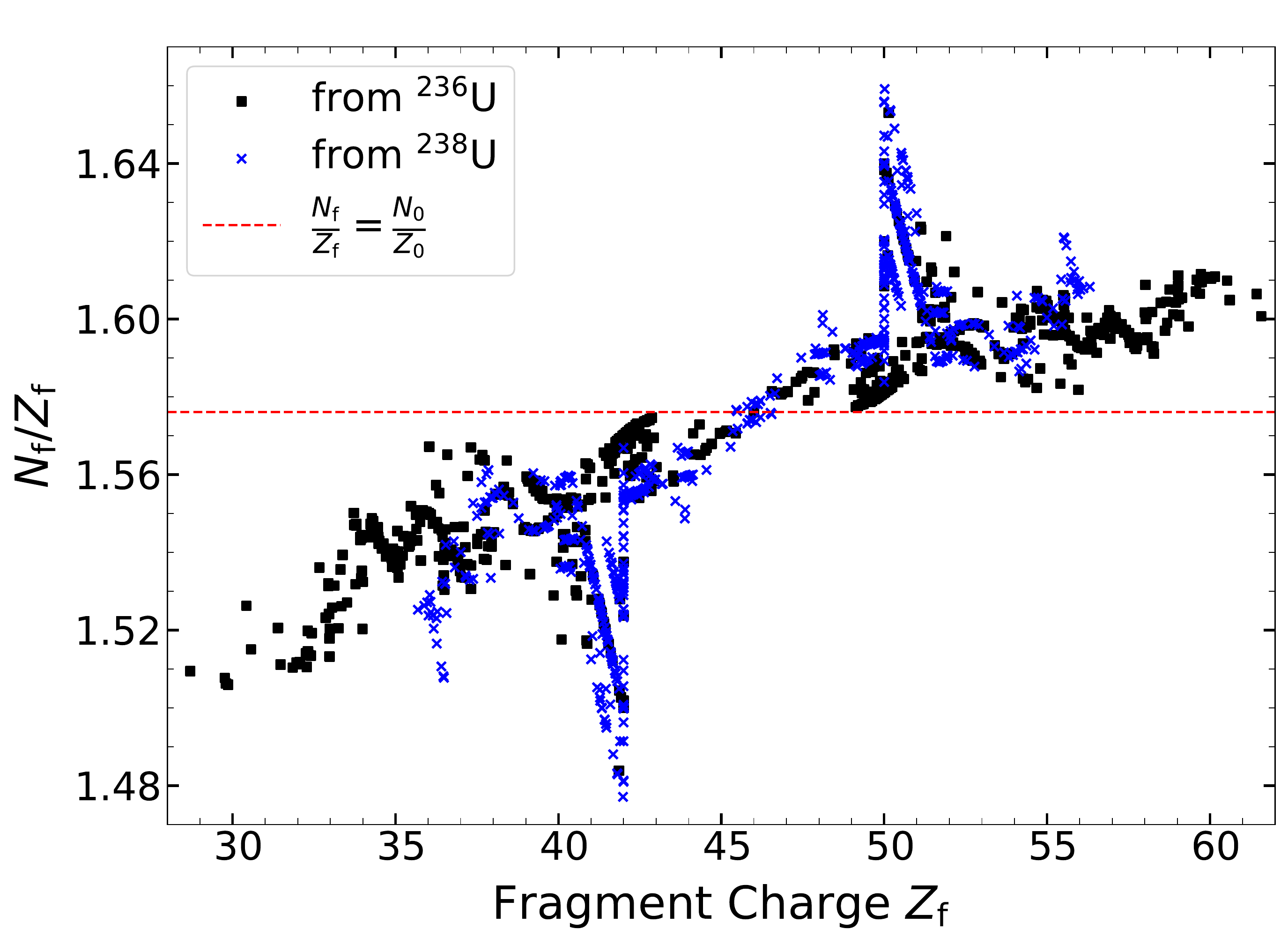}
\caption{Ratio of neutron over proton numbers in the fission fragments of
$^{237}$U as a function of the charge number of the fragment for $K=1/2$. Black 
squares represent blocking calculations initialized  from the PES in $^{236}$U 
and blue crosses represent the ones initialized from the PES in $^{238}$U. The 
dashed line corresponds to the ratio $N_0/Z_0 = 145/92$ of the fissioning 
nucleus.}
\label{fig:u237_NoverZ}
\end{center}
\end{figure}

These differences are minor, as illustrated in Figs.~\ref{fig:u237_Qlm_frag} 
and \ref{fig:u237_NoverZ}. Figure \ref{fig:u237_Qlm_frag} shows the axial
quadrupole and octupole deformation $\beta_2$ and $\beta_3$ of the fission
fragments as a function of their mass. The deformations are defined from the
multipole moments as $\beta_{\lambda} = 4\pi / (3AR^{\lambda}) Q_{\lambda 0}$
with $R = 1.2 A^{1/3}$. Black squares correspond to blocking calculations 
initialized from the PES of $^{236}$U, blue crosses to calculations initialized 
from the PES of $^{238}$U. To increase statistics, we have retained all the 
configurations such that $1 \leq \qneck\leq 8$ that have at least one of their 
nearest neighbors with $\qneck < 1$. The ensemble of all such points give a 
very conservative estimate of the scission region. Overall, there are 
relatively few differences between both sets of deformations. As expected, we 
find that both the heavy and the light fragments are octupole-deformed because 
of the competition between spherical and octupole shell effects 
\cite{scamps2018impact}. This confirms earlier studies of fission in even-even 
actinide nuclei \cite{dubray2008structure,younes2009microscopica}. We also note 
the presence of very deformed fragments around $A_{\rm f} \approx 
\numrange{125}{145}$: such configurations are located near $\quadru\approx 400$ 
b and $\octu\approx \numrange{25}{30}$ b$^{3/2}$ in a region plagued by 
discontinuities in the PES. We do not consider them truly physical.

Figure \ref{fig:u237_NoverZ} completes this picture by showing the ratio of the
number of neutrons to the number of protons in the fragments as a function of 
the charge number of said fragment. The dashed line represents the same ratio 
in the fissioning nucleus, $N_0/Z_0 = 145/92$. Results clearly show a charge 
polarization in the fission fragments, that is, the average number of neutrons 
deviates quite significantly from the Unchanged Charge Distribution (UCD) 
approximation, which postulates that $N_{\rm f} / Z_{\rm f} = N_0 / Z_0$. This 
justifies microscopically the empirical models used to simulate the charge 
polarization \cite{wahl1988nuclearcharge}.

\section{Fission Fragment Distributions}
\label{sec:fpy}

This section summarizes our results on the fission fragment charge and mass 
distributions of the \isotope[236,238]{U}(n,f) reactions. Primary fission 
fragment distributions are extracted from the solution to the TDGCM+GOA 
equation \eqref{eq:tdgcmgoa} of \secref{subsec:lacm}. However, in the case of 
an odd-mass system, the application of the blocking procedure leads to a 
multisheet PES -- one sheet for each $K=\ev*{\Jz}$. We discuss how to set up 
the TDGCM+GOA equation in such a case and how to combine calculations for 
different $K$ values to extract the yields. We then use the code FREYA 
\cite{verbeke2015fission,verbeke2018fission} to model the deexcitation of the 
fission fragments and calculate the independent fission fragment mass and 
charge distributions.

\subsection{Initial Fission Fragment Distributions}
\label{subsec:fpy_ini}

Fission fragment distributions are extracted from the flux of the collective
wave packet solution to Eq.~\eqref{eq:tdgcmgoa} according to the general 
procedure described in detail in Ref.~\cite{regnier2018felix2}. However, a few
additional steps are needed to account for the fact that the compound nucleus
can have different spin projections and that the probability of occupation of
each of these configurations is given by the characteristics in the entrance
channel.

In Sec.~\ref{subsubsec:blocking}, we denoted by $V_K(\qvec)$ the potential 
energy surface for the spin projection $K$ of the odd nucleus. Since the 
collective nuclear Hamiltonian is rotationally invariant, we can compute the 
time-evolution using the TDGCM+GOA equation of motion for each $K$ 
independently, that is,
\begin{equation}\label{eq:tdgcmgoaK}
  i\hbar \pdv{\GCMstate<K>(\qvec,t)}{t}
  = \left[
      \Hcoll<K>(\qvec)
    + i\Acoll<K>(\qvec)
    \right] \,
    \GCMstate<K>(\qvec,t).
\end{equation}
To infer the fission fragment distributions from the set of 
$\GCMstate<K>(\qvec,t)$, we need to determine the initial probability that the
compound nucleus is populated with spin projection $K$. In addition, solving
Eq.~\eqref{eq:tdgcmgoaK} requires setting the initial state for the time 
evolution, i.e., $\GCMstate<K>(\qvec,t=0)$.

\subsubsection{Initial Conditions}

The initial probability $P_{\rm th.}(J^{\pi})$ to populate a given total
angular momentum $J$ and parity $\pi$ is determined using the coupled channel
code FRESCO~\cite{thompson1988coupled} which is part of the LLNL-developed
Hauser-Feshbach code YAHFC (version 3.67) \cite{ormand2021monte}. The set of
rotational states, potentials, and deformation parameters needed to define the
coupled channels calculation were taken from Ref.~\cite{Soukhovitskii:16}; see
Sec. \ref{subsec:entrance}. The probability $p(J^\pi, K)$ to populate each $K$ 
is determined using the equidistribution of the probabilities
\begin{equation}
  p(J^{\pi}, K) = \frac{P_{\rm th.}(J^{\pi})}{2J+1}.
\end{equation}
The probability of populating a given $K$ is the sum of the probabilities for
all valid $J$ and $\pi$. Since we have $-J \leq K \leq J$, we get
\begin{equation}
p(K) = \sum_{J=|K|}^{J_{\rm max}} \frac{P_{\rm th.}(J^{-})+P_{\rm th.}(J^{+})}{2J+1}\, .
\label{eq:probaJtoK}
\end{equation}
In principle, $p(K)$ should be obtained for $J_{\rm max}\rightarrow+\infty$. In
practice, we use a truncation of $J_{\rm max}=\frac{33}{2}$, which is high
enough to obtain a good approximation of the error associated with the other 
truncation in $K$. With this expression, we find that $p(K)=p(-K)$.

\begin{figure}[!ht]
\begin{center}
\includegraphics[width=\columnwidth]{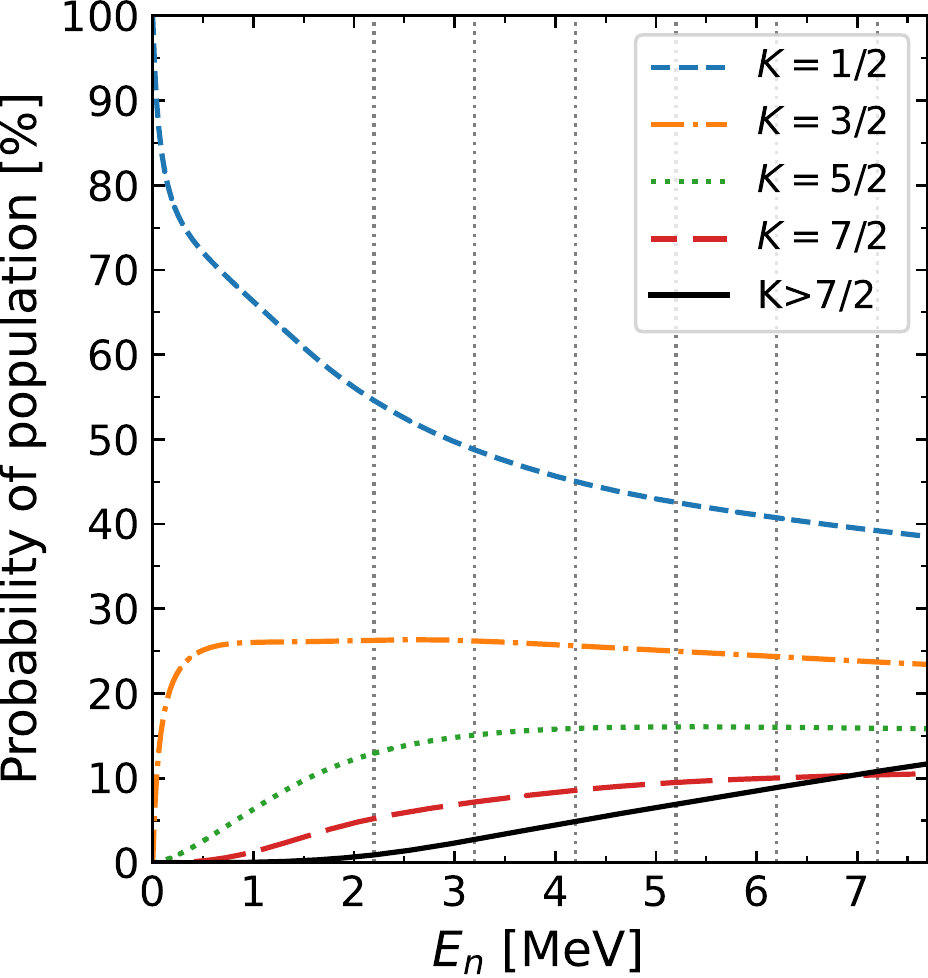}
\caption{\label{fig:u239_initialKpop} Initial probability $p_{\pm}(K)$ to
populate a given spin projection $K$ in the \isotope[238]{U}(n,f) reaction as a 
function of the incident neutron energy $E_n$. Vertical dashed lines represent 
the energies considered in this work.}
\end{center}
\end{figure}

The nuclear Hamiltonian $\Hnucl$ is time-reversal invariant. Consequently, we
get the same time-evolution and the same associated fission yields on a 
potential energy surface with values of $K$ that differ by only a sign. Thus, 
we determine the time-evolution only for $K>0$ with the population probability
$p_{\pm}(K) = p(K) + p(-K)$ given by
\begin{equation}
p_{\pm}(K) = \sum_{J=|K|}^{J_{\rm max}}
\frac{P_{\rm th.}(J^{-})+P_{\rm th.}(J^{+})}{J+\frac{1}{2}}\,.
\end{equation}
The probability $p_{\pm}(K)$ we obtain with our approach is presented in 
\figref{fig:u239_initialKpop} for the $\isotope[238]{U}$(n,f) reaction. We see 
that the contribution from configurations associated with $K>7/2$ is always 
below 11\% for the six neutron energies $\En$ considered here.

\begin{figure}[!ht]
\begin{center}
\includegraphics[width=\columnwidth]{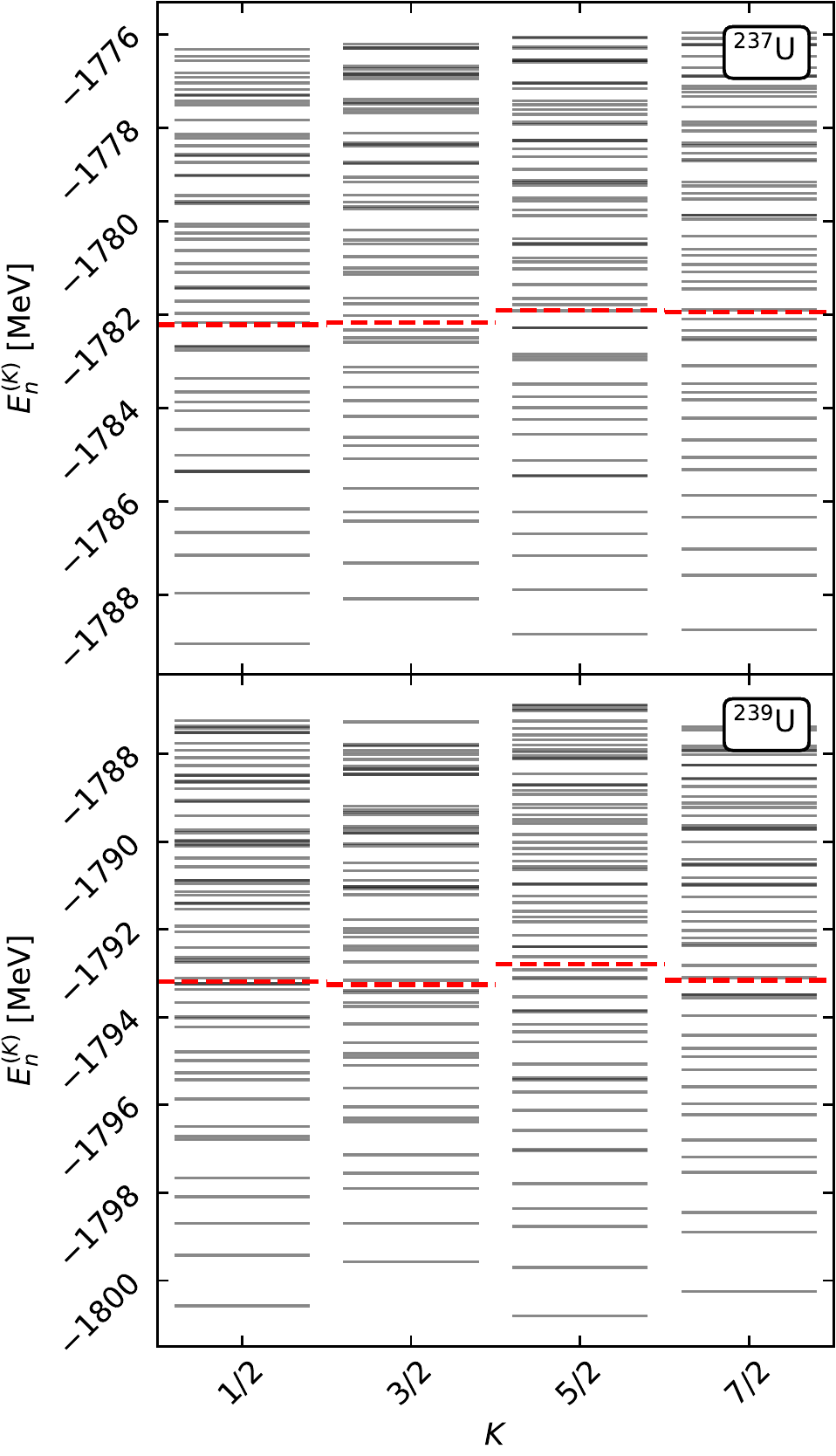}
\caption{\label{fig:spectrumQuasiBound} Spectrum of quasi-bound states for each 
spin projection $K$ obtained by solving the static GCM+GOA equation within the 
extrapolated ground-state potential well; see Ref.~\cite{regnier2018felix2} for 
details. The red dashed line corresponds to the energy at the saddle point, 
which defines the barrier.}
\end{center}
\end{figure}

We define the initial state for the TDGCM+GOA time-evolution for each $K$ using
the prescription of Ref.~\cite{regnier2018felix2}. We recall that this consists
in first determining quasi-bound states $\GCMstateInit$ located in the
ground-state potential well, defined as solutions of the static GCM+GOA
equation in an extrapolated potential, and then build the initial state as a
superposition of these states, where the weights of the superposition are 
Gaussian functions of the energy
\begin{equation}
\GCMstate<K>(\qvec,t=0) = \sum_n
\exp[\frac{\big(\GCMEne<K>_n - \GCMEneAvg<K>\big)^2}{2\GCMSigInit^2}]
\GCMstateInit(\qvec) \,,
\end{equation}
where $\GCMEne<K>_n$ is the energy of $\GCMstateInit$. We show in
\figref{fig:spectrumQuasiBound} the spectra of such quasi-bound states for
\isotope[237,239]{U}.

The width $\GCMSigInit$ is a model parameter that controls the spread of the
initial collective wave packet. In this work, we set $\GCMSigInit=0.5$ MeV. We
determined the level density of quasi-bound states to be approximately 4 
MeV$^{-1}$ at the energy of the barrier and about 7 MeV$^{-1}$ at 5 MeV above 
the fission barrier. Thus, we have between 20 and 40 quasi-bound states 
contributing to the initial state. The parameter $\GCMEneAvg<K>$ is adjusted 
iteratively in order to ensure that the energy of the initial collective wave 
packet matches the physical energy $\GCMEneInit$ of the compound nucleus. It is 
convenient to write
\begin{equation}
\GCMEneInit = \EBbind + E_{\rm x}\,,
\end{equation}
where $\EBbind$ corresponds to the minimum of the saddle point energies over
all $K$,
\begin{equation}
\EBbind = \min_{K}\left(\EBbind<K>\right)\,,
\end{equation}
and $E_{\rm x} = 0, 1, 2, 3, 4, 5 \text{ MeV}$ represents the excitation
energy with respect to this minimum saddle configuration.

\subsubsection{Calculation of the Collective Flux}

We have simulated the large-amplitude collective motion of the fission process
all the way to the formation of the fragments with the computer code 
FELIX~\cite{regnier2018felix2}. For each PES with spin projection $K$, the
absorption field $\Acoll \equiv \Acoll<K>$ in Eq.~\eqref{eq:tdgcmgoaK} is
parameterized by the absorption rate $r$ and width $w$, which is equivalent to
\begin{equation}
\Acoll(\qvec) = \frac{4r}{w^{3}} x^{3}(\qvec)\,,
\end{equation}
where $x(\qvec)$ is zero if $\qvec$ corresponds to a non-scissioned 
configuration, and is equal to the Euclidean distance to the scission line 
otherwise. We fix the ratio $\tfrac{4r}{w^{3}}=0.04$ MeV. The scission line is 
defined as an isoline of the expectation value $\qneck$ of the Gaussian neck 
operator. In this work, we fixed $\qnecksciss=6.5$.

We use the collective inertia tensor defined in \secref{subsubsec:tdgm}. The
zero-point energy correction is extracted from the GCM+GOA width and the
inertia tensor through $\varepsilon = \tfrac{1}{2}\bf{\Gamma}\bf{M^{-1}}$ in
the perturbative cranking approximation of the GCM 
\cite{schunck2016microscopic}. The collective wavefunction $\GCMstate<K>$ is 
discretized using a rectangular cell mesh with a finite element basis of degree 
4, where the nodes are located on the zeros of the Gauss-Lobatto quadrature of 
order 5. We use a timestep of $\Delta t=2.10^{-25}$ s and run the simulation up 
to $t_{\rm max} = 3.10^{-20}$ s.

We model the probability $P_{\rm R}(A, \xi)$ for the right fragment at point
$\xi$ of the scission line to have mass $A$ as an integrated Gaussian,
\begin{equation}
P_{\rm R}(A, \xi) = \int_{A-\frac{1}{2}}^{A+\frac{1}{2}}
\frac{\dd{a}}{\sigmaARes\sqrt{2\pi}}
\exp[ -\frac{\left(a-\AR(\xi)\right)^2}{2\sigmaARes^2} ]\,,
\end{equation}
where $\AR(\xi)$ is the average number of particles in the right fragment and
$\sigmaARes$ is a parameter of our model that represents the particle-number
dispersion in the right fragment and a mass resolution of the experimental data
we use to compare with our results. Following earlier studies
\cite{verriere2021microscopic}, we use $\sigmaARes=4.0$. We then determine the
integrated flux $F(\xi)$ across the scission line $\mathcal{S}$, and determine
the primary fission fragment mass distributions $Y(A)$ through
\begin{equation}
Y(A)= \int_{\mathcal{S}} \dd{\xi} F(\xi)\,P_{\rm R}(A, \xi)\,.
\end{equation}

Finally, we noticed that the determination of the fission fragments by
integration of the flux across the scission line could include spurious
negative contributions caused by a part of the wave packet going back through
the scission line from the opposite direction. We quantify this effect with the
quantity
\begin{equation}
C_{\rm flux}(T)
= \frac%
{\displaystyle\int_0^{T} \dd{t} \int_{\mathcal{S}} \dd{\xi} \max(-\varphi(\xi, t), 0)}%
{\displaystyle\int_0^{T} \dd{t} \int_{\mathcal{S}} \dd{\xi} |\varphi(\xi, t)|}\,.
\end{equation}
The results for the reactions \isotope[236,238]{U}(n,f) are collected in
Table~\ref{tab:negativeflux} for the scission configurations defined by the
condition $\qnecksciss=6.5$.
\begin{table}[!htb]
  \caption{\label{tab:negativeflux} Values of $C_{\rm
  flux}(T\to\infty)$  for the two
  reactions \isotope[236,238]{U}(n,f). All values are given in percent.}
  \begin{ruledtabular}
    \begin{tabular}{c|c|cccccc}
              &     &
          \multicolumn{6}{c}{$E_{\rm x}$ [MeV]} \\
      Target & $K$ &  0 &  1  &  2  &  3 &  4 &  5 \\
      \hline
      \multirow{ 4}{*}{\isotope[236]{U}}%
        & 1/2 & 23.9 & 27.4 & 25.9 & 26.0 & 26.2 & 25.1 \\
        & 3/2 & 23.9 & 24.2 & 23.0 & 23.5 & 22.9 & 22.5 \\
        & 5/2 & 18.1 & 20.2 & 16.2 & 14.2 & 12.4 & 13.8 \\
        & 7/2 & 15.7 & 13.5 & 14.3 & 16.6 & 19.5 & 18.3 \\
      \hline
      \multirow{ 4}{*}{\isotope[238]{U}}%
        & 1/2 & 21.5 & 25.0 & 22.9 & 21.8 & 20.3 & 20.9 \\
        & 3/2 & 17.9 & 19.0 & 19.5 & 19.0 & 18.5 & 20.8 \\
        & 5/2 &  9.3 &  8.5 & 10.2 & 10.4 & 13.6 & 13.6 \\
        & 7/2 & 21.2 & 21.8 & 23.3 & 23.5 & 24.3 & 23.7
    \end{tabular}
  \end{ruledtabular}
\end{table}

We must also associate the different values of $E_{\rm x}$ with the energy of
an incoming neutron  in order to be able to compare our results with 
experimental data. For fissionable isotopes, fission only occurs when the
incident neutron energy is higher than some threshold $\En^{\rm f}$. 
Measurements suggest $\En^{\rm f} \approx 0.7$ MeV for the 
\isotope[236]{U}(n,f) reaction \cite{tovesson2014fast} while $\En^{\rm f} 
\approx 1.2$ MeV for the \isotope[238]{U}(n,f) reaction \cite{tovesson2014fast,
shcherbakov2002neutroninduced}. In such cases, we can assume that our results 
at $E_{\rm x}=0$ should be compared with $E_n = \En^{\rm f}$, and this leads to 
the simple generalization at higher incident energies: $E_n = \En^{\rm f} + 
E_{\rm x}$. One of the limitations in this work is that we assume axial 
symmetry: as mentioned in the introduction to this section, fission barriers 
are therefore systematically overestimated. We account for this effect by 
assuming a generic offset $\Delta E_{\rm triax.}=1$ MeV. This leads to the 
approximate conversion between incident neutron energy and collective energy,
\begin{equation}
\En = \En^{\rm f} + \Delta E_{\rm triax.} + E_{\rm x}.
\label{eq:Ex_map_En}
\end{equation}

Reactions such as \isotope[235]{U}(n,f) and \isotope[237]{U}(n,f) do not have
an energy threshold $\left(\En^{\rm f} = 0\right)$: fission occurs even for 
thermal neutrons \cite{brown2018endf,mcnally1974neutroninduced}. Strictly 
speaking, it is possible that after absorbing a neutron, the compound nucleus 
is at an excitation energy that is substantially higher than the top of the 
highest fission barrier. In Eq.~\eqref{eq:Ex_map_En}, this would correspond, 
effectively, to negative values of $\En^{\rm f}$. Currently, there is no 
theoretical approach sufficiently accurate and precise to verify this 
hypothesis. For fissile nuclei, it is thus best to keep the simple 
relationship: $\En = \Delta E_{\rm triax.}  + E_{\rm x}$.

\subsubsection{Analysis of $^{238}\text{U(n,f)}$}

\begin{figure}[!ht]
\begin{center}
\includegraphics[width=\linewidth]{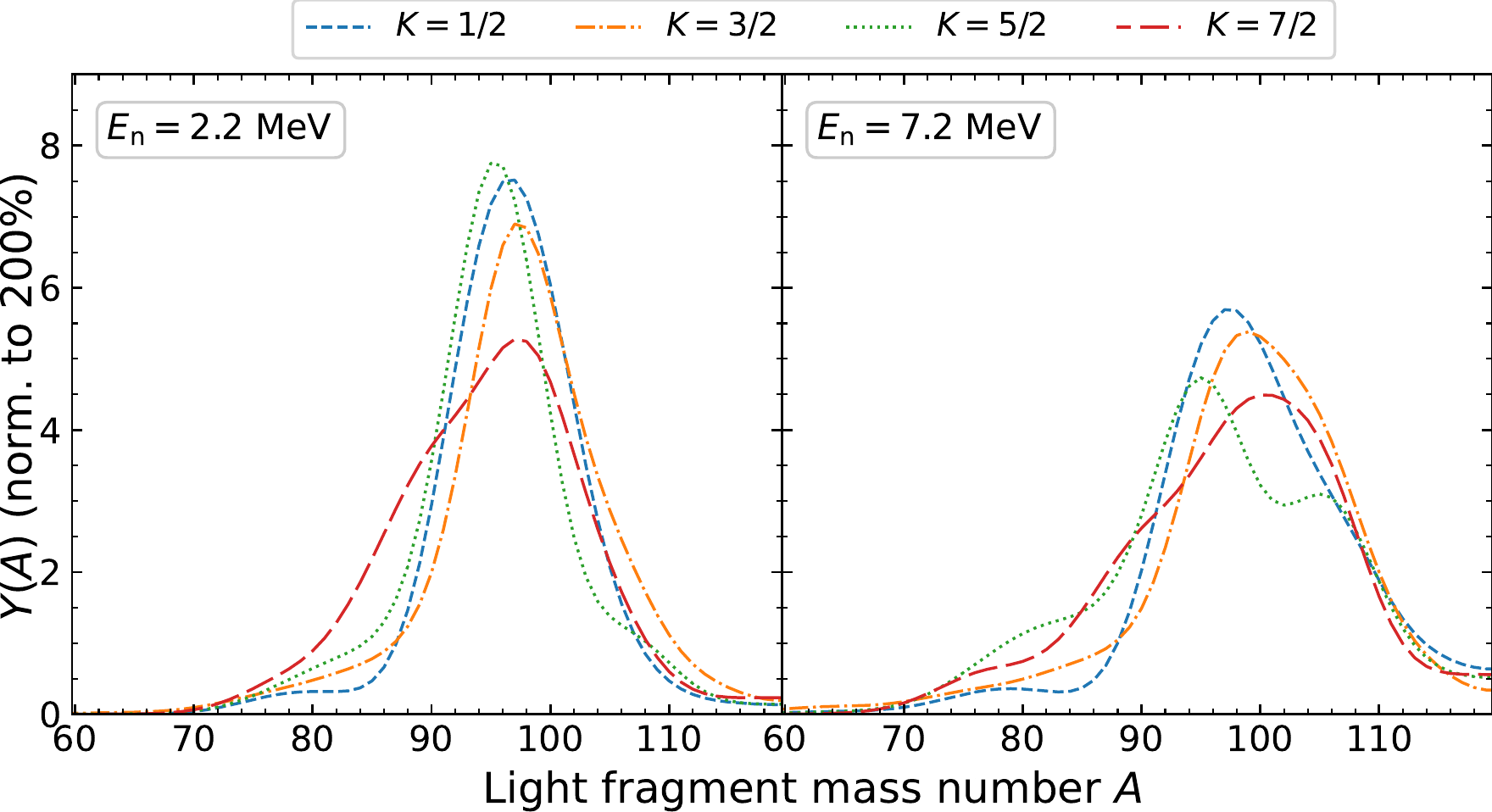}
\caption{\label{fig:u239_primary_yields_K} The first four $K$-components of
the mass distribution of the light fission fragment in the 
\isotope[238]{U}(n,f) reaction, before prompt emission, at incident neutron 
energy $E_n = 2.2$ MeV (left) and $E_n = 7.2$ MeV (right). Each curve has been
independently normalized to 200\%.}
\end{center}
\end{figure}

In this section, we focus on the case of the \isotope[238]{U}(n,f) reaction to 
analyze the impact of the prescription outlined in the two previous sections on 
the primary fission fragment distributions. Figure 
\ref{fig:u239_primary_yields_K} shows the primary mass distributions of the 
light fragment produced in the fission of \isotope[239]{U} for different spin 
projections $K$ and at two different incident neutron energies. The differences 
between the curves for each $K$ value are meaningful since all the ingredients 
in the calculation (definition of the scission configurations, characteristics 
of the TDGCM+GOA, post-processing of the collective flux, etc.) are identical
in all four cases: the only differences are the values of the potential energy 
and collective inertia tensor. Although \figref{fig:u239_initialKpop} shows the 
population probability of $K = \tfrac{1}{2}$ is about twice that of 
$K = \tfrac{3}{2}$, the probability of populating states with higher $K$ is not 
negligible and so the differences in mass distributions are important. As the
incident neutron energy increases, the probability to populate states of higher 
$K$ also increases, further magnifying the importance of considering the 
contributions of different spin projections.

To study the impact of these specific fission-spin channels on the primary mass 
distribution $Y(A)$ we look for the available experimental data. Mass 
distributions from fission reactions induced by fast neutrons are limited and 
only available for some standard fission reactions important for nuclear 
technology. One example is the primary mass distributions of the 
\isotope[238]{U}(n,f) reaction from $E_{n}^{\rm (exp)}=1.2$ to 5.8 
MeV~\cite{vives2000investigation}. The energies of the two fragments after 
prompt particle emission were measured with a dual Frisch-grid ionization 
chamber and the primary fragment masses were determined using the 
double-kinetic energy technique. Provisional masses of the two fragments were 
estimated based on conservation of momentum and the assumption that the 
fragments were detected back-to-back. Then, the energy of the primary fragments 
(before neutron emission) was computed based on the expected number of neutrons 
emitted by each fragment $\nu(A)$. The provisional masses were updated based on 
the pre-neutron energies, and this was repeated until the change in the 
fragment masses from one iteration to the next was less than some fraction of a 
mass unit. The authors of Ref.~\cite{vives2000investigation} assumed a 
sawtooth-like shape for $\nu(A)$; however, the shape of the neutron 
distribution as well as the average total number of emitted neutrons as a 
function of incident neutron energy have been estimated based on available data 
from neighboring fissioning systems. Since the number of neutrons emitted by 
each fragment is unknown, the mass of the fragments in a single event cannot be
determined more accurately than 4-5 mass units (FWHM). It should also be noted 
that using this technique the primary mass yields in light and heavy groups are 
symmetric relative to half of the mass number of the fissioning nucleus 
($A = 239$ in this case).

\begin{figure}[!ht]
\begin{center}
\includegraphics[width=\linewidth]{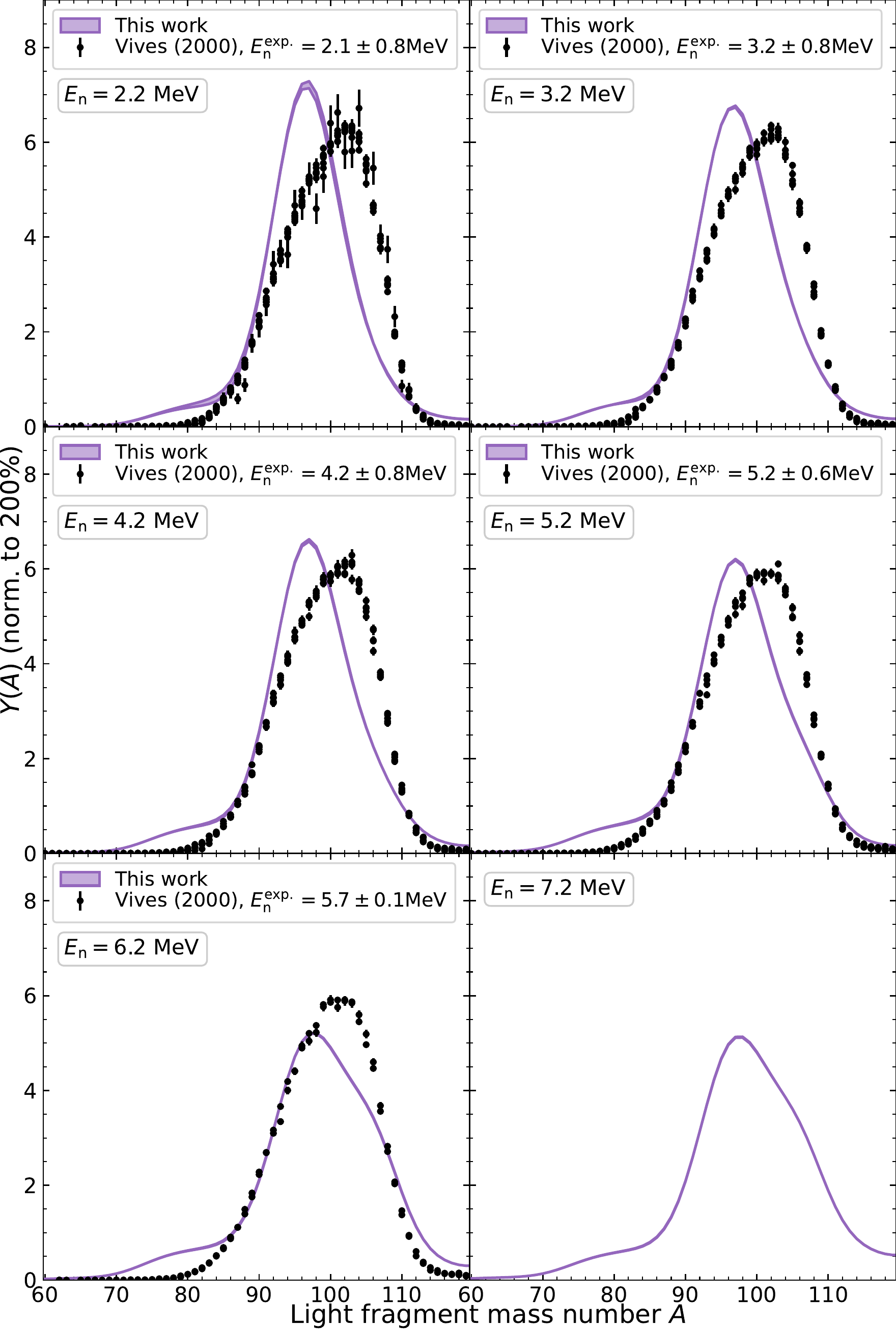}
\caption{\label{fig:u239_primary_yields} Mass distribution of the light fission
fragment in the \isotope[238]{U}(n,f) reaction, before prompt emission, as a
function of incident neutron energy. Experimental data are taken from 
Ref.~\cite{vives2000investigation}.}
\end{center}
\end{figure}

In Fig. \ref{fig:u239_primary_yields} we compare our calculation of these 
primary mass distributions of \isotope[239]{U} (for the light fragment) with 
the available experimental dataset of the \isotope[238]{U}(n,f) 
reaction~\cite{vives2000investigation} for several incident neutron energies up 
to the onset of second-chance fission. The error band was obtained by 
considering an error of $\pm 1$ MeV in the relation given by 
Eq.~\eqref{eq:Ex_map_En}. Overall, the comparison is rather satisfactory for a 
``first-principles'' approach to the calculation of the mass distribution, 
especially since potential energy surfaces in two-dimensional $(\quadru,\octu)$ 
spaces are known to exhibit several spurious discontinuities 
\cite{dubray2012numerical}, the removal of which would require increasing the 
number of collective variables \cite{zdeb2021description,han2021scission,
lau2022smoothing}. In addition, it was also shown that, at least in two 
dimensions, calculations with collective variables based on the expectation 
value of multipole moments could not map all possible fragmentations 
\cite{younes2012fragment}.

\subsection{Independent Yields}
\label{subsec:fpy_ind}

As mentioned in Section \ref{sec:theory}, the primary fission fragments will be 
sufficiently excited to evaporate neutrons in less than 10$^{-15}$ s. These 
very short times mean that in any experiment the nuclei that are detected are 
not the primary fragments, but instead the secondary fragments that have 
emitted a varying number of neutrons. As discussed in the previous section, 
the ``experimental'' primary yields presented in 
Fig.~\ref{fig:u239_primary_yields} were reconstructed from measurements of 
independent yields following a model-dependent procedure.

However, independent yields can also be computed from the primary ones by 
simulating the emission of prompt neutrons and photons. As is commonly known, 
the main drawback of doing so is that one needs to completely characterize the 
fission fragments at scission: not just their distribution $Y(Z,A)$ but also 
their excitation energy $E^{*}$, spin-parity distribution $p(J^{\pi})$, and 
level density $\rho(E^{*},J^{\pi}$). In spite of very encouraging progress in 
recent years, a predictive model of {\it all} such quantities does not yet 
exist  \cite{schunck2022theory}. Evaluations of fission product yields 
typically rely on various empirical models with parameters adjusted to data. We 
adopt a similar strategy here: the prompt emission of particles is simulated 
with the event generator FREYA \cite{verbeke2015fission,verbeke2018fission}.

By default, FREYA can calculate fission events of the \isotope[238]{U}(n,f) 
reaction; various model parameters have already been adjusted to reproduce 
experimental data. Therefore, we used the default FREYA setup to process our 
\isotope[238]{U}(n,f) primary yields with only two modifications: (i) we 
replaced the default 5-Gaussian parameterization of the primary mass 
distribution with our calculated ones at $E_n=2.2$ MeV; and (ii) we changed the 
parameter $dTKE$, which is an overall energy shift to the total kinetic energy. 
FREYA determines the total kinetic energy for a pair of fragments using 
experimental data, and the shift $dTKE$ is tuned to reproduce the prompt 
neutron multiplicity, $\bar{\nu}$. We adjusted $dTKE$ for $E_n=2.2$ MeV from 
its default value of 1.0 MeV to 0.698 MeV in order to match the ENDF/B-VIII.0 
value $\bar{\nu}=2.605$.

\begin{figure}[!ht]
\begin{center}
\includegraphics[width=\columnwidth]{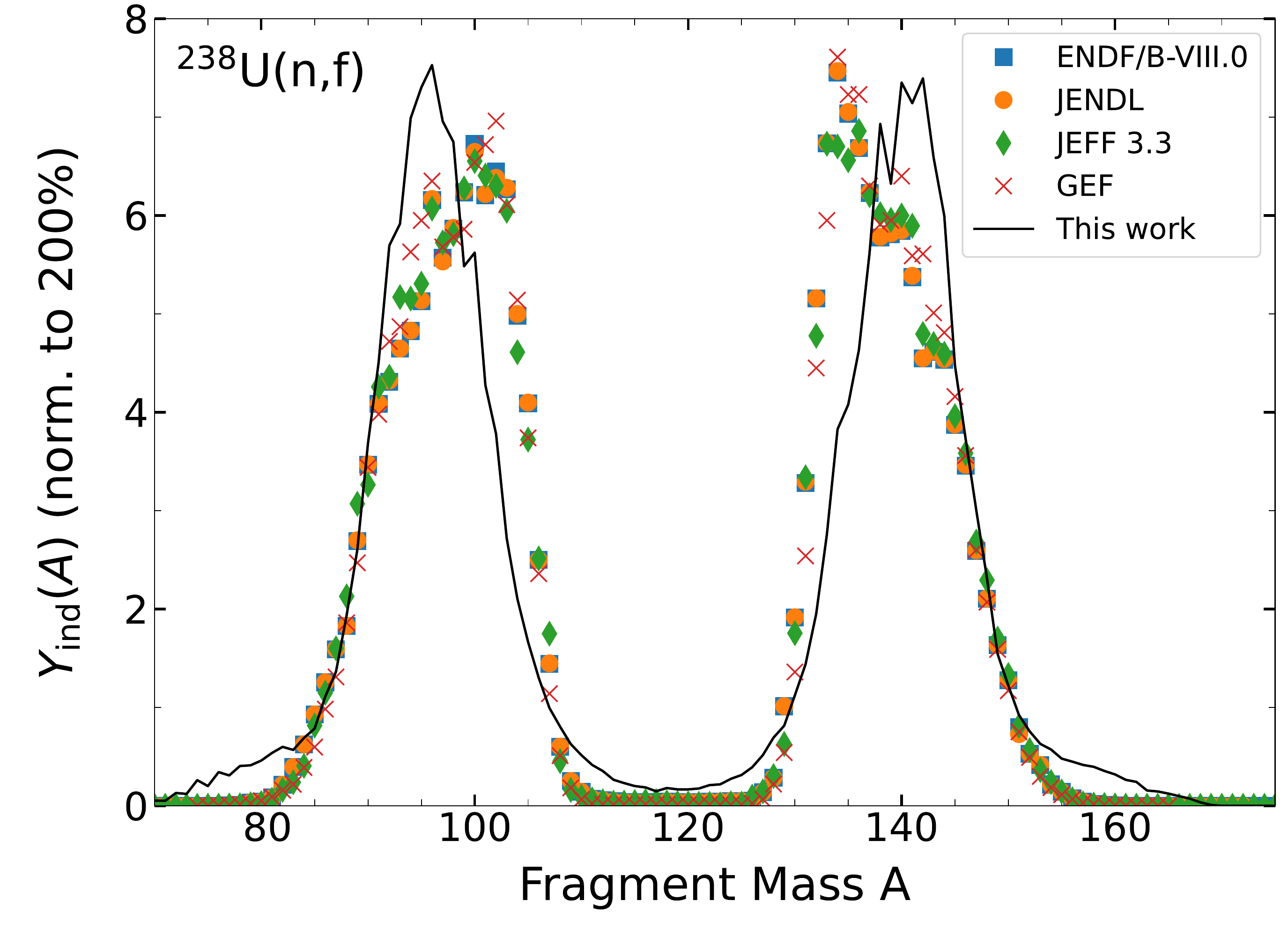}
\caption{\label{fig:u239_independentYA} Fission fragment mass distribution of 
the \isotope[238]{U}(n,f) reaction after neutron emission for an incident 
neutron energy $E_n=$2.2 MeV. The present results are compared with a 
GEF-2021/1.1 \cite{schmidt2016general} empirical model calculation at $E_n=$2.2 
MeV and the following evaluations for fast-neutron-induced fission: 
ENDF/B-VIII.0 \cite{brown2018endf},  JEFF-3.3 \cite{Plompen:2020aa}, and 
JENDL-5 \cite{jendl5}.}
\end{center}
\end{figure}

Figure \ref{fig:u239_independentYA} compares our calculations with several 
evaluations of the independent mass yields in the reaction 
\isotope[238]{U}(n,f) at an incident neutron energy of $E_n = 2.2$ MeV. The 
agreement with the data is rather good considering that the primary mass 
distribution comes from a model prediction rather than an empirical fit. The 
main limitation is that the distance between the two peaks is slightly too 
wide. This is most likely caused by the fact that the heavy fragments in our 
set of scission configurations near the most likely fission is slightly too 
heavy (and the corresponding light fragment slightly too light). We also note 
that symmetric and very asymmetric fission are overestimated. 

FREYA has not been tuned for the \isotope[236]{U}(n,f) reaction, so we added 
that reaction and generally kept the default values of any of the model 
parameters. As for the case of \isotope[238]{U}(n,f), we replaced the default 
primary mass yields with our calculated fission fragment mass distribution, 
this time at $E_n=1.7$ MeV. Pre-equilibrium neutron emission was disabled since 
there is no available data for this process for this reaction. There is also no 
suitable experimental database or evaluation for the total kinetic energy as a 
function of fragment mass for the \isotope[236]{U}(n,f) reaction. For this 
reason, we took the experimental data from the \isotope[235]{U}(n,f) reaction 
instead. The parameter $dTKE$ was set to -1.480 MeV to reproduce the 
ENDF/B-VIII.0 value $\bar{\nu} = 2.545$ for $E_n=$1.7 MeV.
 
\begin{figure}[!ht]
\begin{center}
\includegraphics[width=\columnwidth]{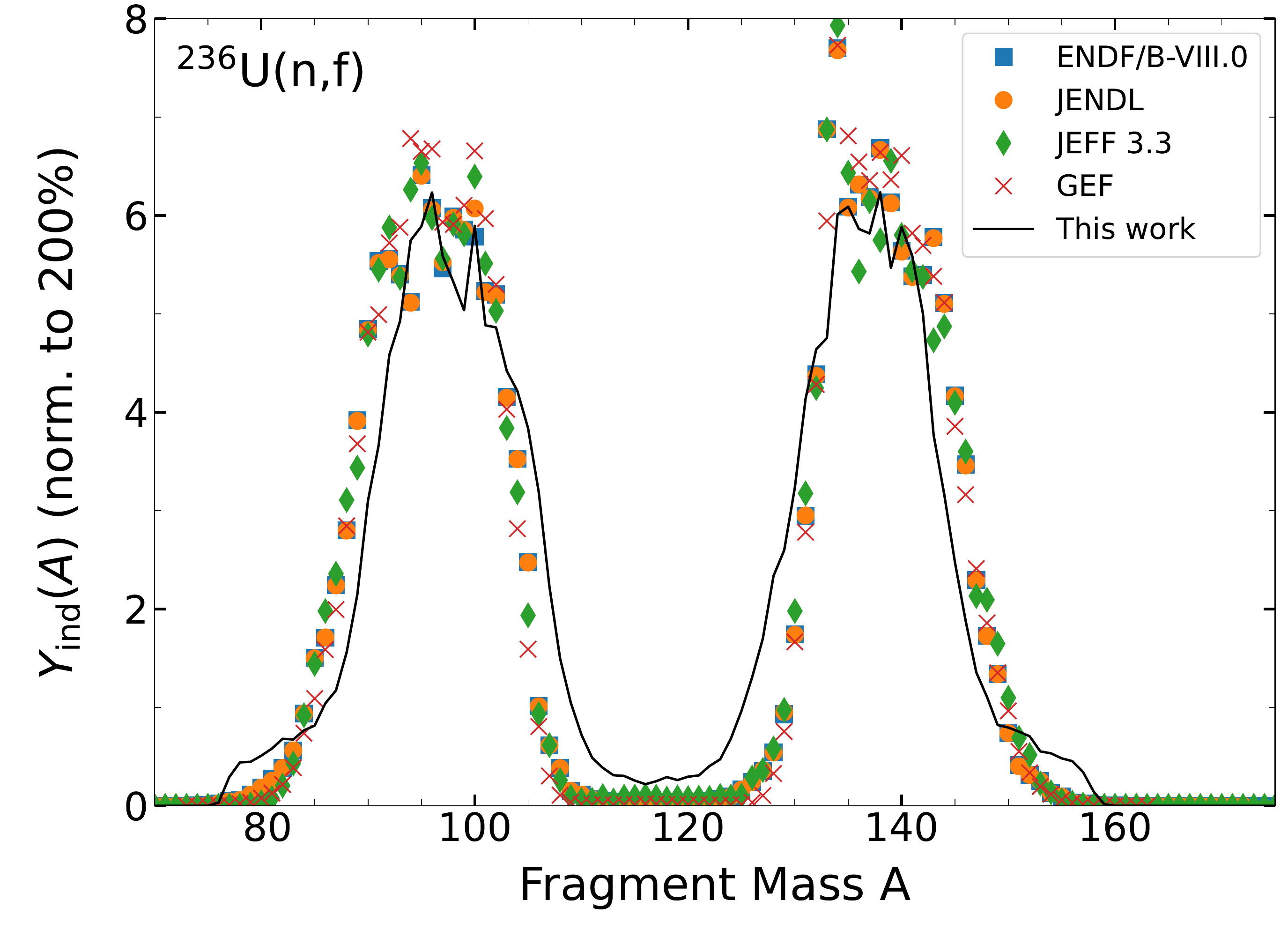}
\caption{\label{fig:u237_independentYA} Fission fragment mass distribution of 
the \isotope[236]{U}(n,f) reaction after neutron emission for an incident 
neutron energy $E_n=$1.7 MeV. The present results are compared with a 
GEF-2021/1.1 \cite{schmidt2016general} empirical model calculation at $E_n=$1.7 
MeV and the following evaluations for fast-neutron-induced fission: 
ENDF/B-VIII.0 \cite{brown2018endf}, JEFF-3.3 \cite{Plompen:2020aa}, and 
JENDL-5 \cite{jendl5}.   }
\end{center}
\end{figure}

As shown in Fig.~\ref{fig:u237_independentYA}, results for 
\isotope[236]{U}(n,f) are somewhat similar to \isotope[238]{U}(n,f). Again, 
both symmetric and very asymmetric fission are overestimated. This time, 
however, the centroids of the light and heavy mass peaks are much closer to the 
evaluated values. The yields we compute are slightly lower than the evaluated 
one, especially in the heavy peak. This may caused by the fact that our 
scission configurations near the most likely fission lack some fragmentations 
around $A_{\rm H} \approx 135$. This problem is reminiscent of issues 
identified earlier in potential energy surfaces obtained with constraints on 
standard multipole moments \cite{younes2012fragment,younes2013microscopic}.

\section{Conclusions}
\label{sec:conclusion}

In this work, we established a rigorous procedure to compute the fission
fragment mass distributions before the emission of prompt neutrons within the
general framework of nuclear energy density functional theory. Our method
assumes that the nuclear shape is axially symmetric and requires three
ingredients: (i) the spin distribution of the fissioning nucleus, which we
obtain from the coupled-channel formalism; (ii) the potential energy surfaces
for different spin projections $K$, which are computed within the
Hartree-Fock-Bogoliubov theory with the equal filling approximation of the
blocking prescription; and (iii) the collective inertia tensor determined by 
the finite-temperature extension of the adiabatic time-dependent
Hartree-Fock-Bogoliubov theory. For the latter, we sketched the complete
derivation of the formulas used without proof so far in the literature.

We tested our approach on the \isotope[236]{U}(n,f) and \isotope[238]{U}(n,f) 
fission reactions, which leads to the odd-mass compound nuclei \isotope[237]{U} 
and \isotope[239]{U}, respectively. We confirmed that the choice of the 
blocking configuration has a major impact on deformation properties: fission
barrier heights, which are key ingredients in the evaluation of fission cross 
sections and probabilities, can vary by up to 1--2 MeV depending on the choice 
of blocked quasiparticle~\cite{koh2017fission}. We also showed that the fission
fragment distributions obtained for different $K$ configurations are
significantly dissimilar and that the different population probabilities of 
each spin channel can magnify these differences. We emphasized that mapping the 
collective wavepacket's energy with the incident neutron's kinetic energy is 
much more challenging in odd-mass systems since each spin channel has a 
different barrier height. We simulated the prompt emission of particles with 
the code FREYA to compare our calculations with experimental data. Overall, the
agreement between our model and experimental data is satisfactory.

Combining our microscopic approach of computing primary fission observables 
with the fission simulation model FREYA opens up the possibility to study the 
impact of different entrance channels on fission-fragment mass distributions. A 
systematic comparison of fission-product yields emerging from different angular 
momenta transfer to the compound system is now potentially feasible. We have 
also provided a framework for large-scale fission calculations of odd-mass 
nuclei involved, e.g., in $r$-process nucleosynthesis, for which no 
experimental data exists. As often in self-consistent calculations, our
two-dimensional potential energy surfaces are plagued by several 
discontinuities. To turn our theoretical framework into a competitive 
evaluation tool, one must enlarge the collective space and develop algorithms 
capable of eliminating spurious discontinuities. Only then will a proper 
quantification of theoretical uncertainties associated with, e.g., the 
definition of scission configurations or the choice of the inertia tensor or 
zero-point energy contributions, will truly make sense.

\begin{acknowledgments}
This work was performed under the auspices of the U.S.\ Department of Energy by
Lawrence Livermore National Laboratory under Contract DE-AC52-07NA27344 and in 
part by the Office of LDRD.
Computing support came from the Lawrence Livermore National Laboratory (LLNL)
Institutional Computing Grand Challenge program.
\end{acknowledgments}


\providecommand{\noopsort}[1]{}

\end{document}